\documentclass[lettersize,journal]{IEEEtran}
\usepackage{amsmath,amsfonts}
\usepackage{mathtools}
\usepackage{steinmetz}
\usepackage{algorithmic}
\usepackage{algorithm}
\usepackage{array}
\usepackage[caption=false,font=normalsize,labelfont=sf,textfont=sf]{subfig}
\usepackage{makecell}

\usepackage{textcomp}
\usepackage{stfloats}
\usepackage{url}
\usepackage{verbatim}
\usepackage{graphicx}
\usepackage{cite}
\usepackage{xcolor}
\usepackage{hyperref}
\hypersetup{hidelinks=true}
\newtheorem{theorem}{Theorem}
\newtheorem{definition}{Definition}
\newtheorem{lemma}{Lemma}
\newtheorem{assumption}{Assumption}

\newcommand{\qed}{\hfill \ensuremath{\Box}}
\usepackage{tikz}
\usetikzlibrary{arrows, positioning, calc}
\usepackage{environ}
\newcommand*\squeezespaces[1]{
  \thickmuskip=\scalemuskip{\thickmuskip}{#1}
  \medmuskip=\scalemuskip{\medmuskip}{#1}
  \thinmuskip=\scalemuskip{\thinmuskip}{#1}
  \nulldelimiterspace=#1\nulldelimiterspace    
  \scriptspace=#1\scriptspace                  
}
\newcommand*\scalemuskip[2]{
  \muexpr #1*\numexpr\dimexpr#2pt\relax\relax/65536\relax
} 

\makeatletter
\newsavebox{\measure@tikzpicture}
\NewEnviron{scaletikzpicturetowidth}[1]{%
	\def\tikz@width{#1}%
	\def\tikzscale{1}\begin{lrbox}{\measure@tikzpicture}%
		\BODY
	\end{lrbox}%
	\pgfmathparse{#1/\wd\measure@tikzpicture}%
	\edef\tikzscale{\pgfmathresult}%
	\BODY
    }
\begin{document}

\title{Frequency Domain Design of a Reset-Based Filter: An Add-On Nonlinear Filter for Industrial Motion Control}

\author{S. Ali Hosseini, Fabian R. Quinten, Luke F. van Eijk, Dragan Kosti\'c, and S. Hassan HosseinNia,~\IEEEmembership{Senior Member~IEEE}

\thanks{This project is co-financed by ASMPT and Holland High Tech, Topsector High Tech Systems and Materials, with a PPS innovation grant public-private collaboration for research and development.}
\thanks{S. Ali Hosseini, Fabian R. Quinten, and S. Hassan HosseinNia are with the Department of Precision and Microsystems Engineering, Delft University of Technology, 2628 CD Delft, The Netherlands (e-mail: S.A.Hosseini@tudelft.nl; F.R.Quinten@student.tudelft.nl; S.H.HosseinNiaKani@tudelft.nl).}
\thanks{Luke F. van Eijk is with ASMPT, 6641 TL Beuningen, The Netherlands, and also with the Department of Precision and Microsystems Engineering, Delft University of Technology, 2628 CD Delft, The Netherlands
(e-mail: luke.van.eijk@asmpt.com).}
\thanks{Dragan Kosti\'c is with ASMPT, 6641 TL Beuningen, The Netherlands (e-mail: dragan.kostic@asmpt.com).}
\thanks{Manuscript received April 19, 2021; revised August 16, 2021.}}
\markboth{Journal of \LaTeX\ Class Files,~Vol.~14, No.~8, August~2021}%
{Hosseini \MakeLowercase{\textit{et al.}}: Closed-loop Frequency Domain Reset Control Design: Application to an Industrial Motion Platform}

\maketitle
\begin{abstract}
This study introduces a modified version of the Constant-in-Gain, Lead-in-Phase (CgLp) filter, which incorporates a feedthrough term in the First-Order Reset Element (FORE) to reduce the undesirable nonlinearities and achieve an almost constant gain across all frequencies. A backward calculation approach is proposed to derive the additional parameter introduced by the feedthrough term, enabling designers to easily tune the filter to generate the required phase. The paper also presents an add-on filter structure that can enhance the performance of an existing LTI controller without altering its robustness margins. A sensitivity improvement indicator is proposed to guide the tuning process, enabling designers to visualize the improvements in closed-loop performance. The proposed methodology is demonstrated through a case study of an industrial wire bonder machine, showcasing its effectiveness in addressing low-frequency vibrations and improving overall control performance.
\end{abstract}


\section{Introduction}\label{Sec: Introduction}
\IEEEPARstart{L}{inear}-Time-Invariant (LTI) controllers are among the most commonly used controllers in industrial applications. Their popularity stems from their ease of tuning and implementation, as well as their ability to be represented in the frequency domain for both open-loop and closed-loop configurations. This characteristic is particularly advantageous because it eliminates the need for a parametric model of the system; instead, the design and tuning can be performed using only the measured frequency response function (FRF) of the system.

In precision positioning platforms, there is a demand for extremely fast and accurate motions, which necessitates pushing controllers to their performance limits. However, LTI controllers face inherent limitations that prevent precision positioning systems from achieving the required levels of speed and accuracy. Bode's gain-phase relationship \cite{freudenberg2000surveyBodeGain}, and waterbed effect \cite{waterbed2012} are two well-known limitations of LTI control systems. 

Over the years, various nonlinear and hybrid control strategies have been proposed to address these limitations, such as variable-gain integrators \cite{hunnekens2014synthesisVariableGain}, hybrid integrator-gain systems \cite{deenen2017hybrid,hosseini2024higher}, and reset control systems \cite{banos2012reset,guo2015analysis}. Among these approaches, reset control systems have demonstrated significant potential for integration into precision motion control systems. Numerous studies have shown that these systems can overcome the limitations of LTI controllers \cite{guo2009frequency,caporale2024practical}. Moreover, reset control systems are advantageous because their open-loop response can be analyzed in the frequency domain using the describing function (DF) method \cite{khalil2002nonlinear}, and closed-loop frequency domain analysis tools are also available \cite{saikumar2021loop,dastjerdi2022closed}. Specifically, \cite{LukeMIMOHOSIDF} introduces higher-order sinusoidal-input describing functions (HOSIDFs) for open-loop reset control systems, while \cite{saikumar2021loop, ZHANG2024106063} introduce the closed-loop HOSIDF for such systems. Building on these results, \cite{dastjerdi2022closed} presents a sensitivity-like function, termed "pseudo-sensitivity," that facilitates closed-loop analysis of reset control systems.

The concept of reset control was first introduced in \cite{clegg1958nonlinear} as a nonlinear integrator, later termed the Clegg integrator (CI). Over time, more sophisticated reset elements were developed, including the First-Order Reset Element (FORE) \cite{horowitz1975nonFORE}, Generalized FORE (GFORE) \cite{guo2009frequency}, and the Second-Order Reset Element (SORE) \cite{hazeleger2016second}.

In \cite{saikumar2019constant}, a reset-based filter, the Constant-in-Gain, Lead-in-Phase (CgLp) element, was introduced. By combining a GFORE element with a lead filter, this design provides broadband phase lead while maintaining an almost constant gain in its describing function. Typically, the CgLp filter is employed to introduce a positive phase near the bandwidth frequency, thereby flexing the limitations imposed by Bode's gain-phase relationship. However, in the CgLp filter, the nonlinear integrator action persists at all frequencies (after the cut-off frequency of the GFORE element) due to the presence of the GFORE element. This results in undesirable nonlinearities at frequencies where such behavior is unnecessary. Moreover, the filter exhibits an inherent low-pass characteristic at high frequencies, which contradicts its intended function of maintaining an almost constant gain.
To address these challenges, this study proposes the following:
\begin{itemize}
    \item Modifying the existing CgLp element by incorporating the feedthrough term in the GFORE element, resulting in a new CgLp element with an almost constant 0 dB gain at all frequencies while exhibiting a reduced ratio of higher-order harmonics to the first harmonic.
\end{itemize}

As noted, the feedthrough term of the GFORE element in this modified CgLp filter is not zero, introducing an additional parameter to be tuned compared to the original CgLp element. This additional parameter increases the complexity of the design process. Furthermore, since the CgLp filter is primarily a phase-generator filter, the required phase is typically predefined in many applications, necessitating the tuning of the filter's parameters to achieve the desired phase. The inclusion of this extra parameter further complicates the tuning process.
To address this issue, this study proposes the following:
\begin{itemize}
    \item An analytical backward calculation approach that derives the additional parameter of the modified CgLp filter from the required phase. This method provides a direct relationship between the CgLp parameters and the phase it generates, enabling designers to achieve the exact desired phase without added complexity.
\end{itemize}

This article examines an industrial wire bonder machine as a case study. This machine establishes physical connections between chips and their packaging, necessitating fast and precise motion. Its primary challenge stems from low-frequency base-frame vibrations, which require an increase in controller gain at low frequencies to mitigate positioning errors. However, due to Bode's gain-phase relationship and the waterbed effect, increasing gain can reduce the phase margin or violate robustness margins (e.g., the peak of the sensitivity function). Thus, the objective is to design a filter that increases gain at low frequencies without compromising robustness margins.

As noted, most industrial applications rely on LTI controllers, with a strong preference for preserving their characteristics while ensuring sufficient performance. Therefore, to further enhance system performance, it is essential to develop an architecture that functions as an add-on filter within existing control loops, eliminating the need for modifications to the implemented linear controller. To address these objectives, this study introduces:
\begin{itemize}
    \item The design of an add-on filter structure, along with step-by-step tuning guidelines, to enable performance improvement for any linear control system without altering the existing linear controller. Additionally, we propose a sensitivity improvement indicator that allows designers to directly shape and tune the nonlinear controller in closed-loop with respect to sensitivity improvements compared to the linear controller.
\end{itemize}

The remainder of this paper is structured as follows. Section \ref{Sec: Priliminaries} introduces the reset element in the time domain. It then presents the open-loop and closed-loop frequency-domain representations of the reset control system. In Section \ref{Sec: CgLp}, we modify the CgLp element by incorporating a proportional GFORE element and examine the effect of this modification on the influence of higher-order harmonics. Additionally, we establish the relationship between the CgLp parameters and the required phase, enabling the filter to be designed independently from the added feedthrough term (thereby avoiding unnecessary complexity). In Section \ref{sec: case study}, we introduce the wire bonder machine as the case study and highlight the limitations of linear control and existing challenges. Subsequently, in Section \ref{Sec: reset control of the wire bonder}, we present the design and tuning method for a reset-based add-on filter that addresses the challenges observed in the LTI control of the wire bonder, while also providing a general design and tuning method applicable to other systems. Additionally, we introduce the sensitivity improvement indicator, a frequency-domain-based method that visualizes the closed-loop performance comparison and prediction between the linear and reset-based controllers. In Section \ref{sec: experimental result}, we validate the findings of this study through experiments conducted on the industrial wire bonder machine. Finally, conclusions and suggestions for future work are provided in Section \ref{Sec: Conclusion}.

\section{Reset Control Preliminaries}\label{Sec: Priliminaries}
This section covers the fundamentals of reset control, including reset element definition, and frequency domain analysis for both open-loop and closed-loop reset control systems (RCS).
\subsection{Reset Element}\label{SubSec: Reset element}
Consider a closed-loop control system, as depicted in Fig. \ref{Fig: Block diagram CL}, where \( (r(t), d_i(t), d_n(t)) \in \mathbb{R} \) represent the exogenous inputs, \( u(t) \in \mathbb{R} \) is the control input, and \( y(t) \in \mathbb{R} \) denotes the controlled output, all at \( t \in \mathbb{R}_{\geq 0} \). The LTI plant is represented by \( G \), while \( C_1 \) and \( C_2 \) are LTI filters. The reset element is denoted by \( \mathcal{R} \) and is defined as follows:

\begin{equation}
\label{eq: reset state space}
\mathcal{R} : 
\begin{cases}
    \dot{x}_r(t) = A_r x_r(t) + B_r e_r(t), & \text{if } \left(x_r(t), e_r(t)\right) \notin \mathcal{F}, \\[5pt]
    x_r(t^+) = A_\rho x_r(t), & \text{if } \left(x_r(t), e_r(t)\right) \in \mathcal{F}, \\[5pt]
    u_r(t) = C_r x_r(t) + D_r e_r(t), &
\end{cases}
\end{equation}
where the reset surface \( \mathcal{F} \) is given by:
\begin{equation}
\label{reset surface}
\mathcal{F} := \{ e_r(t) = 0 \wedge (A_\rho - I) x_r(t) \neq 0 \},
\end{equation}
with states $x_r(t) \in \mathbb{R}^{n_r\times 1}$, and after reset states $x_r(t^{+})\in\mathbb{R}^{n_r\times 1}$. The state-space matrices of the reset element are given by \( A_r \in \mathbb{R}^{n_r \times n_r} \), \( B_r \in \mathbb{R}^{n_r \times 1} \), \( C_r \in \mathbb{R}^{1 \times n_r} \), and \( D_r \in \mathbb{R} \). The reset value matrix is denoted as \( A_\rho = \text{diag}(\gamma_1, \dots, \gamma_{n_r}) \), where \( -1<\gamma_{i}<1 \, \forall i\in \mathbb{N}\). $e_r(t)\in\mathbb{R}$ and $u_r(t)\in\mathbb{R}$ represent the input and output of the reset element, respectively. A reset element follows its base linear system (BLS) dynamics if no reset happens $\left((x_r(t), e_r(t)) \notin \mathcal{F}\,\,\, \forall\,\,t \in \mathbb{R}_{\geq 0}\right)$. Thus, the transfer function of its BLS is defined as follows
\begin{equation}
\label{eq RCS bls}
R(s) = C_r(sI - A_r)^{-1}B_r + D_r,
\end{equation}
where $s \in\mathbb{C}$ is the Laplace variable.
\usetikzlibrary {arrows.meta}
\tikzstyle{block} = [draw,thick, fill=white, rectangle, minimum height=2em, minimum width=2.5em, anchor=center]
\tikzstyle{sum} = [draw, fill=white, circle, minimum height=0.6em, minimum width=0.6em, anchor=center, inner sep=0pt]
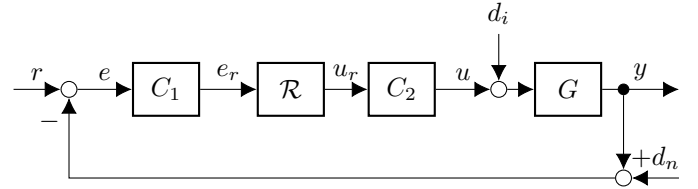
\begin{figure}[!t]
	\centering
	\begin{scaletikzpicturetowidth}{\linewidth}
		\begin{tikzpicture}[scale=\tikzscale]
			\node[coordinate](input) at (0,0) {};
			\node[sum] (sum1) at (1,0) {};
			\node[sum] (sum3) at (8.75,0) {};
			\node[sum] (sum4) at (11,-1.6) {};
			\node[sum, fill=black, minimum size=0.4em] (dot2) at (11,0) {};		
			
			\node[block] (lead) at (5.0,0) {$\mathcal{R}$};
			\node[block] (controller) at (7.0,0) {$C_2$};
			\node[block] (fo-higs) at (2.75,0) {$C_1$};
			\node[block] (system) at (10,0) {$G$};
			\node[coordinate](output) at (12,0) {};
			\node[coordinate](di-input) at (8.75,1) {};
			\node[coordinate](n-input) at (12,-1.6) {};
			\draw[arrows = {-Latex[width=6pt, length=6pt]}] (input)  -- node[above]{$r$} (sum1);
			\draw[arrows = {-Latex[width=6pt, length=6pt]}] (di-input)node[above]{$d_i$}  --  (sum3);
			\draw[arrows = {-Latex[width=6pt, length=6pt]}] (n-input)  -- node[above]{$+d_n$} (sum4);
			\draw[arrows = {-Latex[width=6pt, length=6pt]}] (sum1)   --node[above]{$e$}  (fo-higs);
			\draw[arrows = {-Latex[width=6pt, length=6pt]}] (fo-higs) --node [above]{$e_r$} (lead);
			\draw[arrows = {-Latex[width=6pt, length=6pt]}] (lead)  --node[above] {$u_r$} (controller);
			\draw[arrows = {-Latex[width=6pt, length=6pt]}] (controller)  --node[above] {$u$} (sum3);
			\draw[arrows = {-Latex[width=6pt, length=6pt]}] (sum3) -- (system);
			\draw[arrows = {-Latex[width=6pt, length=6pt]}] (system)  -- node[above]{$y$} (output);
			\draw[arrows = {-Latex[width=6pt, length=6pt]}] (dot2) -- (sum4);
			\draw[arrows = {-Latex[width=6pt, length=6pt]}] (sum4) -| node[pos=0.85,left]{$-$} (sum1);
			
		\end{tikzpicture}
	\end{scaletikzpicturetowidth}
	\caption{Block diagram of the closed-loop system.}
	\label{Fig: Block diagram CL}	
\end{figure}

\subsection{Open-loop Frequency Domain Analysis of Reset Control Systems}\label{subsec: openloop frequency}
The nonlinear nature of reset elements presents significant challenges in designing controllers within the frequency domain, particularly when employing the widely used loop shaping technique, which relies on Bode plots. Among the few methods available to estimate nonlinear controllers in the frequency domain, the describing function method stands out. The describing function characterizes the steady-state response of a convergent nonlinear system by representing it through the first harmonic component of the Fourier series expansion. In \cite{guo2009frequency}, the describing function method is utilized to represent the reset element in the frequency domain. Building upon this, \cite{saikumar2021loop} introduces the extension of the frequency domain tool known as Higher-Order Sinusoidal-Input Describing Functions (HOSIDFs) for reset controllers, enabling a more comprehensive open-loop analysis. Thus, having the input of the open-loop as $e(t)=\hat{e}\sin(\omega t)$, the output $y(t)$ (while the loop is not closed) can be described by the Fourier series:
\begin{equation}
    \label{eq: e to y}
y(t)=\sum_{n=1}^{\infty}\left|\mathcal{L}_n(j\omega)\right|\hat{e}\sin\left(n\omega t +\angle \mathcal{L}_n(j\omega)\right),
\end{equation}
with $n\in \mathbb{N}$, and $\mathcal{L}_n(j\omega)$ is given as follows (see \cite{caporale2024practical,LukeMIMOHOSIDF})
\begin{equation}
    \label{eq: Ln open loop}
\mathcal{L}_n(j\omega)=G(nj\omega)C_2(nj\omega)H_n(\omega)C_1(j\omega)e^{j(n-1)\angle C_1(j\omega)}.
\end{equation}
The $H_n(\omega)$ is the HOSIDF of the reset element $\mathcal{R}$, which is introduced in \cite{saikumar2021loop} as the following theorem.\\

\begin{theorem}
\label{theorem: reset hosidf}
\cite[Theorem 3.1]{saikumar2021loop}  
The HOSIDFs of the reset element in \eqref{eq: reset state space} are expressed as follows:  
\begin{equation} 
H_n(\omega) = 
\begin{cases}
C_r\left(j\omega I - A_r\right)^{-1}\left(I + j\Theta_\text{D}(\omega)\right)B_r+D_r,\\
\hfill 
\text{for } n = 1,\\
C_r\left(jn\omega I - A_r\right)^{-1}j\Theta_\text{D}(\omega)B_r,\\
\hfill
\text{for odd } n \geq 2,\\
0,  
\hfill
\text{for even } n \geq 2,
\end{cases}
\label{eq reset HOSIDF}
\end{equation}  
where the terms are defined as:  
\begin{equation}
\begin{aligned}
&\Lambda(\omega) = \omega^2 I + A_r^2, \\  
&\Delta(\omega) = I + e^{\frac{\pi}{\omega}A_r}, \\  
&\Delta_r(\omega) = I + A_\rho e^{\frac{\pi}{\omega}A_r}, \\  
&\Gamma_r(\omega) = \Delta_r^{-1}(\omega)A_\rho\Delta(\omega)\Lambda^{-1}(\omega), \\  
&\Theta_\text{D}(\omega) = -\frac{2\omega^2}{\pi}\Delta(\omega)\left[\Gamma_r(\omega) - \Lambda^{-1}(\omega)\right].  
\end{aligned}
\end{equation}  
\end{theorem}

Using Theorem \ref{theorem: reset hosidf} and the expression in \eqref{eq: Ln open loop}, the open-loop frequency response of the RCS can be determined. However, due to the presence of a nonlinear element in the loop, closed-loop frequency domain functions, such as sensitivity and complementary sensitivity, do not adhere to the conventional open-loop/closed-loop relationships observed in LTI systems. Consequently, a direct calculation of the closed-loop HOSIDFs is necessary for reset control systems.

\subsection{Closed-loop Frequency Domain Analysis of Reset Control Systems}
As mentioned earlier, unlike LTI systems, there is no direct link between the open-loop and closed-loop frequency domain response for nonlinear controllers, especially in the case of the reset control system. Therefore, predicting the closed-loop performance is desirable using only frequency domain information of the loop components, including the plant. In Theorem \ref{Theorem: Sn theorem}, a frequency domain-based performance prediction for RCSs is presented under the following assumptions \cite{saikumar2021loop}.
\begin{assumption}
    \label{Assumption: convergence}
    \cite[Assumption 1]{saikumar2021loop} The reset control system in Fig. \ref{Fig: Block diagram CL} is input to state convergent.
\end{assumption}
This assumption ensures a unique periodic solution exists for the output of the reset element, which shares the same period as the reset input. Consequently, the reset output signal can be represented using a Fourier series. The validity of this assumption can be evaluated using the FRF-based method presented in \cite[Corollary 1 and Theorem 2]{hosseini2025frequency}.

\begin{assumption}
    \label{Assumption: two reset per cycle}
   \cite[Assumption 3]{saikumar2021loop} Only the first harmonic of $e_r$ results in
resets and hence the creation of higher-order harmonics ($n > 1$) in $u_r$.
\end{assumption}
Assumption \ref{Assumption: two reset per cycle} is crucial for predicting the performance of closed-loop RCSs, as the method proposed in \cite{saikumar2021loop} does not consider the effect of higher-order harmonics on the reset events. However, it has also been demonstrated in \cite{saikumar2021loop} that this assumption holds true when a well-designed reset controller avoids introducing excessively high values for the open-loop HOSIDFs.\\

\begin{theorem}
    \label{Theorem: Sn theorem}
    \cite[Theorem 2]{caporale2024practical}
    Considering $r(t)=\sin{(\omega t)}$ and that Assumption \ref{Assumption: convergence} and \ref{Assumption: two reset per cycle} hold, the closed-loop steady-state error $e_\text{ss}(t)=\lim_{t\rightarrow \infty} e(t)$ can be written as 
    \begin{equation}
    \label{eq ess}
        e_\text{ss}(t)=\sum_{n=1}^{\infty}e_n(t),
    \end{equation}
    where
    \begin{equation}
    \label{eq en}
        e_n(t)=|S_n(j\omega)|\sin{(n\omega t+\angle{S_n(j\omega)})},
    \end{equation}
    with higher-order sinusoidal input sensitivity function $S_n(j\omega)$ as
\begin{equation}
\label{eq Sn}
S_n(j\omega) =
\begin{cases}
\displaystyle \frac{1}{1 + \mathcal{L}_1(j\omega)}, \hfill \text{for } n = 1, \\[10pt]
\displaystyle -\mathcal{L}_n(j\omega) S_\mathrm{bl}(jn\omega) 
\left(|S_1(j\omega)| e^{jn\angle S_1(j\omega)}\right), \\
\hfill \text{for odd } n \geq 2, \\[5pt]
0, \hfill \text{for even } n \geq 2,
\end{cases}
\end{equation}
 where $S_\mathrm{bl}(jn\omega)=\frac{1}{1+L_\mathrm{bl}(jn\omega)}$ which $L_\mathrm{bl}(j\omega)=C_1(j\omega)R(j\omega)C_2(j\omega)G(j\omega)$ is the base linear transfer function of the open-loop.\\
\end{theorem}

Theorem \ref{Theorem: Sn theorem} and the results from \cite{saikumar2021loop} demonstrate the existence of closed-loop HOSIDFs, which must be considered for a more precise design of reset controllers. However, analyzing reset controllers by accounting for all harmonics is a non-trivial task. To facilitate the analysis of reset control systems, in \cite{dastjerdi2022closed}, a method that combines all harmonics into a single frequency function is introduced. This closed-loop approximation, referred to as the pseudo-sensitivity, is defined in \cite[Definition 5]{dastjerdi2022closed} as follows:
\begin{equation}
    \label{eq: S infty}
\left|S_\infty(\omega)\right|=\frac{\max\limits_{0\leq t<2\pi/\omega} e_\text{ss}(\omega,t)}{r_o},
\end{equation}
where $e_\text{ss}(\omega,t)$ can be calculated from \eqref{eq ess} based on HOSIDFs of the closed-loop system excited by $r(t)=r_0\sin(\omega t)$ with $r_0 \in \mathbb{R}_{>0}$. The magnitude \( \left|S_\infty(\omega)\right| \) over the range of frequencies of interest can approximate the sensitivity function of an RCS. While it represents the worst-case scenario, it still facilitates the evaluation of various design effects. Moreover, if \( \left|S_\infty(\omega)\right| \) demonstrates sufficient robustness margins, it serves as a reliable indicator of the system's performance, as it accounts for the maximum steady-state error, \( \max e_\text{ss}(\omega,t) \).

Again, note that a parametric model of the plant is not required to calculate this pseudo-sensitivity, as Theorem \ref{Theorem: Sn theorem} only necessitates the FRF of the plant. This allows for the approximation of the steady-state error based solely on the frequency response of the loop components. Utilizing the tools introduced in this section for designing and analyzing reset control systems in the frequency domain, the following sections will illustrate their applications in the development and evaluation of a reset control system for an industrial motion platform.

\section{Reset-Based Filter Design: A Phase Generator Approach}\label{Sec: CgLp}
Considering the Clegg integrator, characterized by \( A_r = 0 \), \( A_\rho = 0 \), and \( D_r = 0 \) in \eqref{eq: reset state space}, it represents the simplest form of a reset element. Unlike a linear integrator, which introduces a phase lag of \(-90^\circ\), the Clegg integrator exhibits a phase lag of approximately \(-38^\circ\) in its describing function (\( H_1 \) in Theorem \ref{theorem: reset hosidf}) \cite{guo2009frequency}. Although this advantage is evident only in the describing function of reset elements, it can also be utilized to overcome the limitations of LTI controllers \cite{zhao2019overcoming}.

Building on the mentioned phase advantage, a reset-based filter is introduced in \cite{saikumar2019constant} to extend the application of reset control for broadband phase compensation across the desired frequency range. This so-called CgLp filter could replace part of the differentiation action in PID controllers or be used to compensate for the phase lag of any additional filter, as it helps improve the system's precision according to the loop-shaping concept \cite{karbasizadeh2020benefiting, saikumar2019constant}. However, the CgLp element presented in \cite{saikumar2019constant} does not maintain a constant gain across all frequencies. This limitation arises from including a GFORE element combined with a lead-lag filter, which results in a gain slope of \(-20 \, \text{dB/dec}\) at high frequencies. This \(-20 \, \text{dB/dec}\) behavior originates from the nonlinear integrator component of the GFORE element, propagating the nonlinearity to very high frequencies where nonlinear action is unnecessary. In this study, we propose a proportional GFORE-based CgLp filter, where the feedthrough term in the GFORE element is non-zero (\(D_r \neq 0\)), achieving an almost constant gain across all frequencies.
\usetikzlibrary {arrows.meta}
\tikzstyle{sum} = [draw, fill=white, circle, minimum height=0.0em, minimum width=0.0em, anchor=center, inner sep=0pt]
\tikzstyle{block} = [draw,thick, fill=white, rectangle, minimum height=2.5em, minimum width=3em, anchor=center]
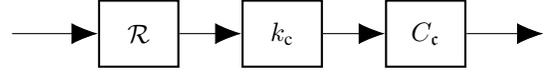
\begin{figure}[!t]
	\centering
	\begin{scaletikzpicturetowidth}{0.8\linewidth}
		\begin{tikzpicture}[scale=\tikzscale]
			\node[coordinate](input) at (0,0) {};
			
			\node[sum] (sum3) at (6.3,0) {};
			

			\node[block] (lead) at (3.2,0) {${k_\text{c}}$};
			\node[block] (controller) at (4.9,0) {$C_\mathfrak{c}$};
			\node[block] (fo-higs) at (1.5,0) {$\mathcal{R}$};
			
			\node[coordinate](output) at (7.2,0) {};

			\draw[arrows = {-Latex[width=8pt, length=10pt]}] (input)  -- node[above]{} (fo-higs);
			\draw[arrows = {-Latex[width=8pt, length=10pt]}] (fo-higs) --node [above]{} (lead);
			\draw[arrows = {-Latex[width=8pt, length=10pt]}] (lead)  --node[above] {} (controller);
			\draw[arrows = {-Latex[width=8pt, length=10pt]}] (controller)  --node[above] {} (sum3);
			
			
		\end{tikzpicture}
	\end{scaletikzpicturetowidth}
	\caption{The CgLp filter structure.}
	\label{Fig: Block diagram CgLp}	
\end{figure}
\begin{definition}
    \label{Lemma: New CgLp}
    In this study, we define the CgLp filter as illustrated in Fig. \ref{Fig: Block diagram CgLp}, where
     \begin{equation}\label{eq: kc}
        k_\text{c} = \frac{\omega_f - \omega_l}{\omega_f},
    \end{equation}
    \begin{equation}\label{eq: Cc}
        C_\mathfrak{c}(s) = \frac{1 + s / \omega_l}{1 + s / \omega_f},
    \end{equation}
    with \( [\omega_l, \omega_f] \in \mathbb{R}^{1\times 2}_{>0} \), and \( \mathcal{R} \) is a proportional GFORE element ($n_r=1$) characterized by
    \begin{equation} \label{eq: Dr}
         A_r = -\omega_r, \quad B_r = 1, \quad C_r = \omega_r, \quad D_r = \frac{\omega_l}{\omega_f - \omega_l},
    \end{equation}
     with $\omega_r\in \mathbb{R}_{>0}$ as
    \begin{equation}
    \label{eq: corner frequency}
    \omega_r = \frac{\omega_l}{\sqrt{1 + \left(\frac{4(1 - A_\rho)}{\pi(1 + A_\rho)}\right)^2}}.\\
    \end{equation}
\end{definition}

The CgLp filter in Definition \ref{Lemma: New CgLp} exhibits an almost constant gain across all frequencies in its first-order describing function ($\mathfrak{C}_1(\omega)$), where:  
\begin{equation}  
    \label{eq: CgLP HOSIDF}  
    \mathfrak{C}_n(\omega) = k_\text{c} C_\mathfrak{c}(n j \omega) H_n(\omega).  
\end{equation}
In Appendix \ref{Appendix: CgLp lemma}, we prove this characteristic of the CgLp element by demonstrating that \( |\mathfrak{C}_1(\omega)| = 1 \) at both low (\( \omega \to 0 \)) and high (\( \omega \to \infty \)) frequencies. Since \( H_1(\omega) \) represents the describing function of the proportional GFORE element and cannot precisely match the magnitude of its linear counterpart, achieving an exact match across all frequencies is impossible. Consequently, a slight deviation occurs in the mid-frequency range; however, it is negligible with respect to the produced phase, similar to the findings in \cite{saikumar2019constant}.

Furthermore, it has been demonstrated that reset elements exhibit less phase lag compared to their corresponding base linear systems \cite{guo2009frequency}. Consequently, the reset element \( \mathcal{R} \) in the CgLp filter does not entirely cancel the positive phase contribution of the lead filter \( C_\mathfrak{c} \) within the frequency range \( [\omega_l, \omega_f] \). As a result, the CgLp filter, as defined in Definition \ref{Lemma: New CgLp}, introduces a positive phase within this frequency range while maintaining an approximately 0 dB gain across the entire frequency spectrum, as proven in Appendix \ref{Appendix: CgLp lemma}.  

Following Definition \ref{Lemma: New CgLp}, we construct the CgLp filter such that it introduces no change in gain when incorporated into a pre-designed linear controller. Additionally, it increases the phase at the frequencies of interest, thereby enabling modifications to the linear controller that were previously unattainable due to Bode's gain-phase relationship. The primary distinction between this CgLp and the one presented in \cite{saikumar2019constant} lies in the inclusion of a non-zero feedthrough term, \(D_r\). This term suppresses the integral action in the GFORE element beyond a threshold frequency (\(\omega_f\)), resulting in a greater difference between the magnitudes of the first-order (\(\mathfrak{C}_1(\omega)\)) and higher-order harmonics (\(\mathfrak{C}_n(\omega)\)) of the CgLp element. 

Thus, the use of this proportional GFORE element requires a tuning method for $D_r$, as, in the previous implementation of the CgLp with a GFORE element, $D_r$ was set to zero. From Definition \ref{Lemma: New CgLp}, it can be observed that \( D_r \) can be calculated based on \( \omega_l \) and \( \omega_f \). The parameter \( \omega_l \) is set according to the desired corner frequency of the lead element ($C_\mathfrak{c}(j \omega)$), as it is also related to the corner frequency of the DF of the GFORE, $\omega_r$, as well. Depending on the system dynamics and external disturbances, \( \omega_l \) can be tuned accordingly. In \cite{saikumar2019constant}, \( \omega_f \) appears only in \( C_\mathfrak{c} \) and is typically set sufficiently large (\( \omega_f \gg \omega_l \)). However, in this context, since \( \omega_f \) directly influences \( D_r \), we aim to determine its value based on known parameters and the desired phase produced by the CgLp element (\( \theta_\text{CgLp} \)). This method obviates the need for introducing an additional parameter for tuning. In the following lemma, we show that the maximum achievable phase of the CgLp occurs as \( \omega_f \to \infty \). Then, we introduce a theorem in which \( \omega_f \) is calculated based on any required phase of the CgLp filter within the achievable range.\\

\begin{lemma}
\label{lemma: max theta}
Let \(\theta_\text{M}(\omega)\) denote the maximum achievable phase of the CgLp at the frequency \(\omega \in \mathbb{R}_{>0}\), defined as
\begin{equation}
\theta_\text{M}(\omega) =\max\limits_{\omega_f \in [\omega_l, \infty)} \left(\theta_\text{CgLp}(\omega,\omega_f)\right),
\end{equation}
where \(\omega_l\) and \(A_\rho\) are fixed. Then, \(\theta_\text{CgLp}(\omega,\omega_f) = \theta_\text{M}(\omega)\) if and only if \(\omega_f \to \infty\).
\end{lemma}

\textbf{Proof}: 
Having $\mathfrak{C}_1(\omega)$ as the describing function of the CgLp element as below, 
\begin{equation}\label{eq: c1}
\mathfrak{C}_1(\omega)= k_\text{c}H_1(\omega)C_\mathfrak{c}(j\omega),
\end{equation}
for $\theta_\text{CgLp}(\omega,\omega_f)$ we have
\begin{equation}
\label{eq: theta cglp}
     \begin{split}
&\theta_\mathrm{CgLp}(\omega,\omega_f) = \\
&\arctan{\left(\frac{b(\omega)}{a(\omega)+\frac{\omega_l}{\omega_f-\omega_l}}\right)} + \arctan{\left(\frac{\omega}{\omega_l}\right)} - \arctan{\left(\frac{\omega}{\omega_f}\right)},
\end{split}
\end{equation}
with (based on Theorem \ref{theorem: reset hosidf})
\begin{equation}
\begin{split}
    &a(\omega)=\mathfrak{R}\left( C_r\left(j\omega I-A_r\right)^{-1}\left(I+j\Theta_D(\omega)\right)B_r\right), \\
    &b(\omega)=\mathrm{I}\left( C_r\left(j\omega I-A_r\right)^{-1}\left(I+j\Theta_D(\omega)\right)B_r\right),
    \end{split}
\end{equation}
where \(\mathfrak{R}(\cdot)\) denotes the real part and \(\mathrm{I}(\cdot)\) denotes the imaginary part. For simplicity, in the remainder of this paper, we use \( a \) and \( b \) instead of \( a(\omega) \) and \( b(\omega) \), respectively. Moreover, it can be observed that \( a(\omega) \) and \( b(\omega) \) are independent of \(\omega_f\).

Since \(\arctan\) is a monotonically increasing function, the expressions \(\max{\left(\frac{b(\omega)}{a(\omega)+\frac{\omega_l}{\omega_f-\omega_l}}\right)}\) and \(\min{\left(\frac{\omega}{\omega_f}\right)}\) result in the maximum value of \(\theta_\mathrm{CgLp}(\omega,\omega_f)\) in \eqref{eq: theta cglp}. Where by having \(\omega_f \to \infty\), results in
\begin{equation*}
    \frac{\omega}{\omega_f}\rightarrow 0,
\end{equation*}
and
\begin{equation*}
    \left(\frac{b(\omega)}{a(\omega)+\frac{\omega_l}{\omega_f-\omega_l}}\right)\rightarrow \frac{b(\omega)}{a(\omega)}.
\end{equation*}
Thus, for fixed values of \( a(\omega) \) and \( b(\omega) \), \(\theta_\text{CgLp}(\omega,\omega_f) = \theta_\text{M}(\omega)\) if and only if \(\omega_f \to \infty\).
 \qed \\

Based on the maximum achievable phase of the CgLp element, as derived in Lemma \ref{lemma: max theta}, the following theorem calculates the required value of \( \omega_f \) (and consequently \( D_r \)) to achieve the desired phase of the CgLp, \( \theta_\text{CgLp} \in (0, \theta_\text{M}) \).\\

\begin{theorem}
\label{lemma: wf calculation}
Given that $\theta_\text{CgLp}(\omega) \in (0, \theta_\text{M}(\omega))$ represents the desired phase of the DF of the CgLp filter ($\mathfrak{C}_1$), the frequency $\omega_f\in [\omega_l,\infty)$ can be determined as follows (for known $\omega_l$ and $A_\rho$):
\begin{equation}
\omega_f=
\begin{cases}
\min(\omega_{f_1},\omega_{f_2}), & \text{if both} \,\omega_{f_1},\omega_{f_2} \in [\omega_l,\infty),\\
\max(\omega_{f_1},\omega_{f_2}), & \text{otherwise},
\end{cases}
\end{equation}
where
\begin{equation}
    \begin{split}
    \omega_{f_1}&=\frac{-k_2-\sqrt{k_2^2-4k_1k_3}}{2k_1},\\
    \omega_{f_2}&=\frac{-k_2+\sqrt{k_2^2-4k_1k_3}}{2k_1},\\
        k_1&=aQ(\omega)-b, \\
        k_2&=b\omega Q(\omega)+b\omega_l+a\omega-(a-1)\omega_lQ(\omega), \\
        k_3&=-\omega\omega_l(bQ(\omega)+a-1),\\
    \end{split}
\end{equation}
with
\begin{equation}
    Q(\omega)=\tan{\left(\theta_\mathrm{CgLp}(\omega)-\arctan{\left(\frac{\omega}{\omega_l}\right)}\right)}.
\end{equation}

\textbf{Proof}: See Appendix \ref{Appendix: wf calculation}.
\end{theorem}

\begin{figure}[!t]
\centering
\subfloat[]{\includegraphics[width=0.8\columnwidth]{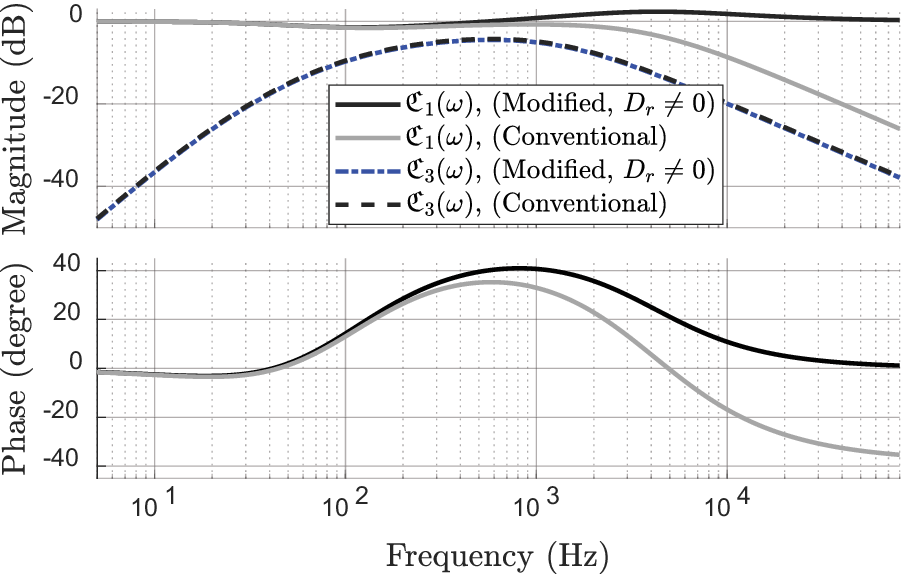}%
\label{Fig: CgLp DF}}
\hfil
\subfloat[]{\includegraphics[width=0.8\columnwidth]{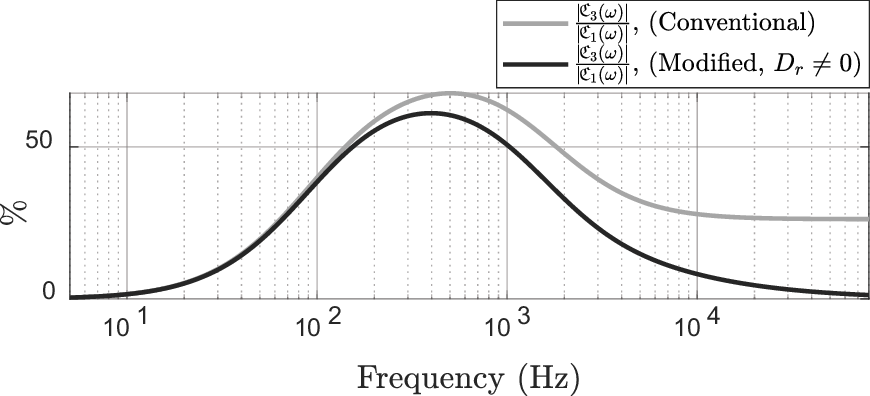}%
\label{fig: C3 over C1}}
\caption{(a) First-order and third-order DF of the CgLp element in both modified and conventional cases with $\omega_l=6.28\times10^{2}\,$rad/sec, $\omega_f=2.51\times10^{4}\,$rad/sec, and $A_\rho=0$ ($D_r$ is set to zero for the conventional case). (b) The relative magnitude between the third- and first-order harmonics of the CgLp elements.}
\label{fig: CgLp}
\end{figure}

As stated in Lemma \ref{lemma: max theta}, the maximum phase of the CgLp filter can be achieved by allowing \(\omega_f \rightarrow \infty\). However, this implies that the nonlinear integrator component of the CgLp must remain active at very high frequencies, which increases the influence of higher-order harmonics in the system. To address this, Theorem \ref{lemma: wf calculation} provides the necessary calculation for \(\omega_f\) to achieve the desired \( \theta_\text{CgLp}(\omega) \) at a specific frequency. This approach enables the designer to mitigate the impact of higher-order harmonics without introducing an additional degree of tuning, which might be undesirable.

In Fig.~\ref{Fig: CgLp DF}, the first-order and third-order describing functions of the CgLp element are depicted for both the conventional and modified CgLp configurations. Both versions share the same parameters; however, in the conventional CgLp, the feed-through term ($D_r$) is set to zero, as in \cite{saikumar2019constant}.

It can be observed that the first-order harmonic in the modified CgLp filter is positioned further away from its third-order harmonic, thereby enhancing reliability in DF-based loop-shaping control design. This effect is further illustrated in Fig.~\ref{fig: C3 over C1}, where the relative magnitude of the third-order describing function over the first-order describing function \(\left(\frac{|\mathfrak{C}_3(\omega)|}{|\mathfrak{C}_1(\omega)|}\right)\) is plotted in percentage for both CgLp cases, clearly demonstrating how the feedthrough term contributes to the reliability of the describing function analysis. Additionally, it can be seen that at high frequencies, in the presence of the feedthrough term, this ratio approaches zero, minimizing the impact of higher-order harmonics on system performance.

Furthermore, it is evident that in this new CgLp filter, the gain remains almost constant not only up to a threshold frequency but extends to \(\omega \rightarrow \infty\). This property ensures that by incorporating this filter into a pre-designed linear controller, the system's loop gain remains consistent while simultaneously improving phase characteristics. This advantage is particularly beneficial in scenarios where a linear controller has already been implemented. In such cases, an add-on filter can be employed, allowing the nonlinear filter to be directly applied without altering the existing linear controller. This approach enhances system performance further while maintaining the integrity of the original control design.

In the next section, we introduce an industrial motion stage where the limitations of the linear controllers leave no room for further improvement. Subsequently, in Section \ref{Sec: reset control of the wire bonder}, we demonstrate how the proposed new CgLp filter effectively addresses the existing control challenges in this industrial setup.

\section{An Industrial Case Study}\label{sec: case study}
This section provides a detailed overview of an industrial wire-bonding machine, outlines the control objectives, introduces the optimal linear controller designed specifically for the system, and discusses the challenges that remain after implementing linear control.
\subsection{Wire Bonding Process}\label{subsec: wire bonding process}
\begin{figure}[!t]
\centering
\subfloat[]{\includegraphics[width=0.25\columnwidth]{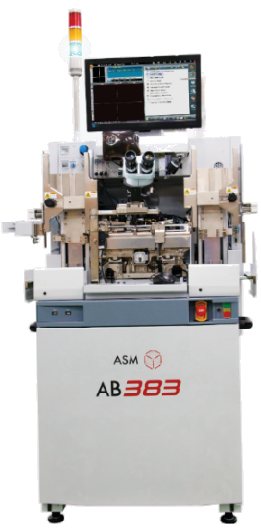}%
\label{fig: AB383}}
\hfil
\subfloat[]{\includegraphics[width=0.75\columnwidth]{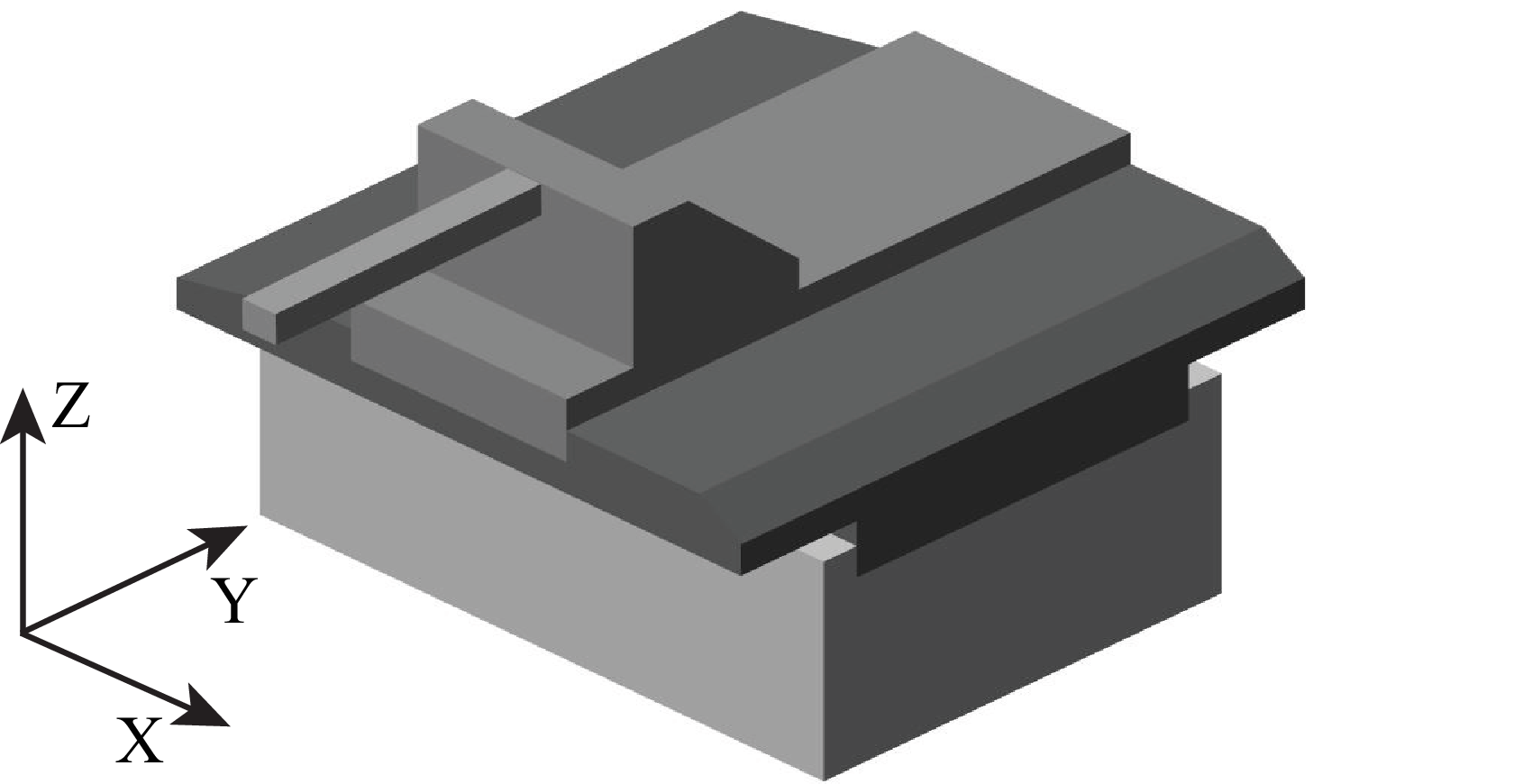}%
\label{fig:simscape model}}
\caption{(a) The industrial wire bonder. (b) Simscape Multibody model of the wire bonder motion platform.}
\label{fig: wirebonder}
\end{figure}

A wire bonder (see Fig. \ref{fig: AB383}) is a machine that connects conducting wires between an integrated circuit and its packaging, forming a microchip. It bonds the wire's ends to their underlying surfaces using thermal or ultrasonic energy. 
 
 The motion platform of the wire bonder possesses three degrees of freedom, which are controlled using a Cartesian coordinate system. To facilitate understanding, a simplified Simscape Multibody model of the wire bonder's motion platform is provided in Fig. \ref{fig:simscape model}. This figure also illustrates the base frame, whose primary function is to isolate vibrations between the motion stages and the external environment. Each axis is equipped with a dedicated actuator that directly applies force to the respective stage. The motion stage is designed and calibrated such that each motion axis can be considered a SISO LTI system within its operational range.

\begin{figure}[!t]
\centering
\includegraphics[width=1\columnwidth]{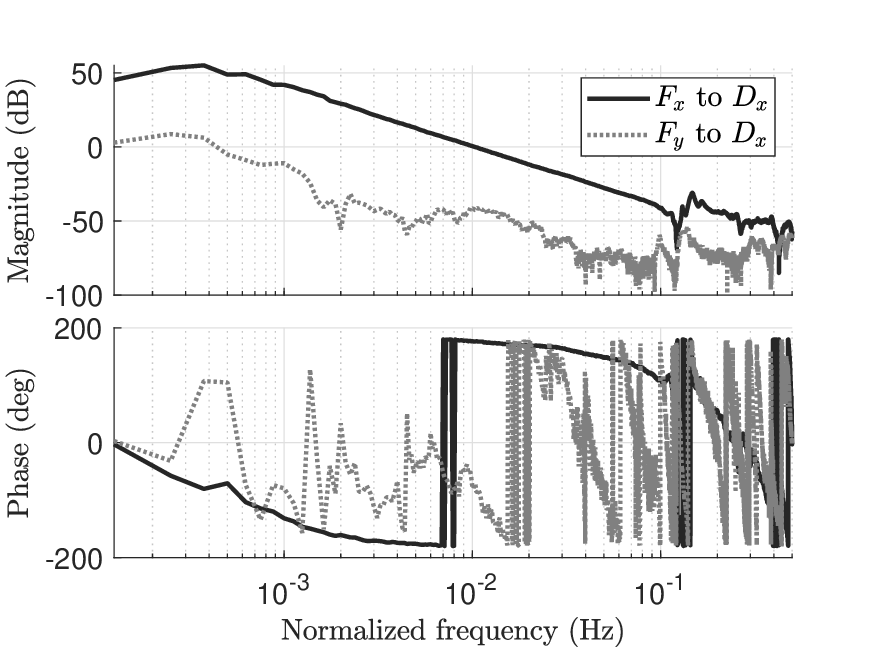}
\caption{FRF of the X-stage of the physical wire bonder, illustrating the mapping of actuator forces in the X-stage ($F_x$) and Y-stage ($F_y$) to the displacement measured by the X-stage encoder ($D_x$).}
\label{Fig: FRF AB383}
\end{figure}

 Fig. \ref{Fig: FRF AB383} illustrates the measured FRF of the physical wire bonder, mapping the generated force \( F_x \), applied to the X-stage, to the resulting displacement \( D_x \). The FRF represents the linearized system, measured around the configuration where all three actuators are positioned at their central locations. Additionally, the figure depicts the FRF from the actuator of the Y-stage to the encoder position of the X-stage. The results indicate that the influence of cross-coupling effects is negligible up to frequencies exceeding the target control bandwidth. Please note that the frequency axis in Fig. \ref{Fig: FRF AB383}, as well as in all other experimental results presented in this study, has been normalized for confidentiality purposes (scaled by an arbitrary factor).

\begin{figure}[!t]
\centering
\includegraphics[width=0.8\columnwidth]{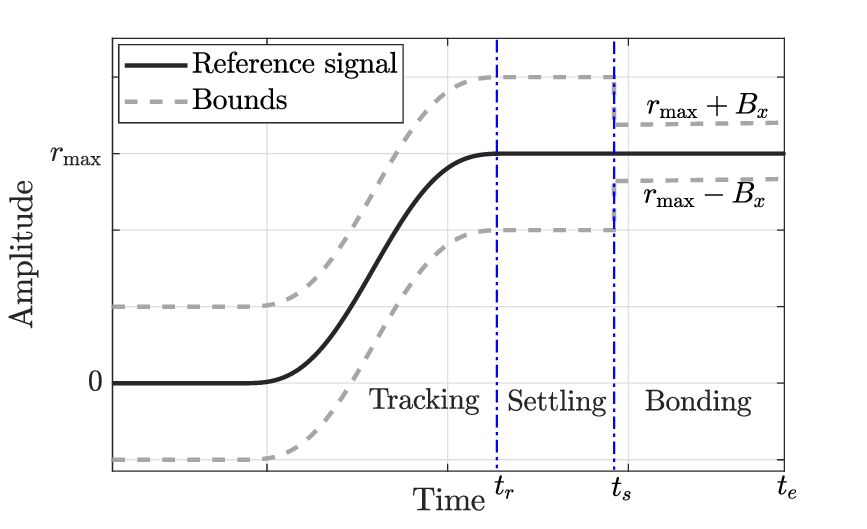}
\caption{Typical reference signal profile in the X-direction, divided into the tracking, settling, and bonding phases. The indicated bounds are exaggerated.}
\label{Fig: Typical Reference}
\end{figure}

Fig. \ref{Fig: Typical Reference} illustrates the general shape of the reference signal used for movement in the X-direction. The movement is divided into three distinct stages. The first stage is the tracking phase, in which the end-effector moves from one bonding surface to the next. The second stage is the settling phase, providing time for the end-effector's oscillations to stabilize within acceptable limits. The final stage, known as the bonding phase, marks the completion of the movement. Surrounding the reference signal are permissible error bounds for the end-effector. These bounds are crucial for preventing contact with previously bonded wires and surrounding objects during the tracking and settling phases. Furthermore, the error bounds during the bonding phase are necessary to ensure the end-effector remains within the bond pads. Consequently, we define the settling time of the system as the last time the position of the end-effector is located outside the $r_\mathrm{max}\pm B_\mathrm{x}$. To give an indication of the machine precision, note that this process enables bonding with an accuracy in the micrometer range. Moreover, it operates at a rate of approximately ten wire bonds per second, highlighting its efficiency

Therefore, a well-designed controller is required to accurately follow the reference signal with minimal error within defined boundaries. The objective is to minimize the settling region while satisfying specified frequency-domain robustness constraints. In the subsequent section, we present the frequency-domain constraints and evaluate the performance of an LTI controller applied to the X-stage of the wire bonder.

\subsection{Linear Control of the Wire Bonder}\label{subsec: linear control for wire bonder}
Industrial motion stages are typically controlled using advanced feedforward controllers designed to follow predefined reference trajectories. Feedback control is primarily employed to enhance tracking accuracy by mitigating disturbances and addressing dynamics not accounted for by the feedforward controller. To achieve high motion control performance, a feedforward control is designed for the industrial wire bonder as well. However, the focus of this study is solely on feedback control, and detailed information about the existing feedforward controller is intentionally omitted.

For a feedback control system with the X-stage wire bonder as the plant (\(G\)), an acceptable LTI controller (\(C_\text{L}\)) in the frequency domain must satisfy the following constraints:
\begin{equation}
    \label{eq: constraints}
    \begin{split}
        &M_\text{s} \leq 6 \, \text{dB}, \\
        &M_\text{r} \leq 2.5 \, \text{dB}, \\
    \end{split}
\end{equation}  
where \(M_\text{s} = \max\limits_{\omega<\omega_\text{res}} |S(j\omega)|\), with \(S(j\omega) = 1/\left(1 + C_\text{L}(j\omega)G(j\omega)\right)\) being the sensitivity function. Additionally, \(M_\text{r} = \max\limits_{\omega \geq \omega_\text{res}} |S(j\omega)|\), where \(\omega_\text{res} \in \mathbb{R}_{>0}\) denotes the frequency prior to the first resonance or anti-resonance of the plant.  
The constraint on \(M_\text{r}\) ensures that even in the presence of inaccuracies in the measured FRF of the plant, the resonance of the system does not become excessively amplified beyond a specified robust bound. Additionally, the upper bound on the maximum magnitude of the sensitivity function ($M_\text{s}$) determines the robustness of the system, as the modulus margin (MM) follows the equality $\text{MM} = \frac{1}{M_\text{s}}$, where $\text{MM}$ denotes the shortest distance between the Nyquist curve of the open-loop FRF and the point $(-1, 0j)$ in the Nyquist diagram. In this study, an automatically tuned LTI controller is considered to push performance to the limits of linear control. The controller aims to achieve optimal tracking performance, maximizing the bandwidth while increasing the loop gain, and simultaneously satisfying the robustness conditions outlined in \eqref{eq: constraints}.

\begin{figure}[!t]
\centering
\includegraphics[width=0.8\columnwidth]{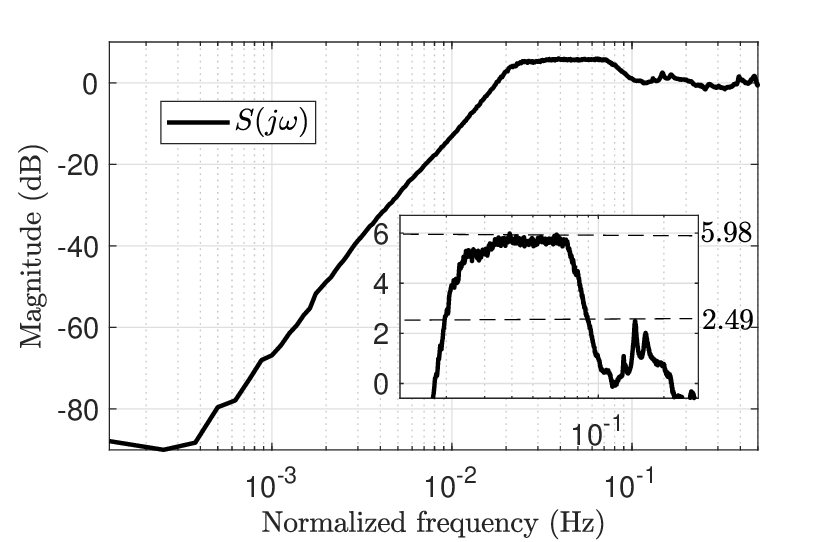}
\caption{Sensitivity of the wire
bonder with the auto-tuned linear controller.}
\label{Fig: S linear}
\end{figure}

Having the $C_\text{L}(j\omega)$ as the automatically tuned discretized linear controller and $G$ as the FRF of the industrial wire bonder, the sensitivity function is plotted in Fig. \ref{Fig: S linear}. It can be observed that the auto-tuner algorithm designed an LTI controller to reduce the magnitude of the sensitivity at low frequencies while maintaining certain robust bounds. The broad range of the sensitivity peak enables a reduction in its magnitude at low frequencies due to the waterbed effect. 

\begin{figure}[!t]
\centering
\subfloat[]{\includegraphics[width=0.8\columnwidth]{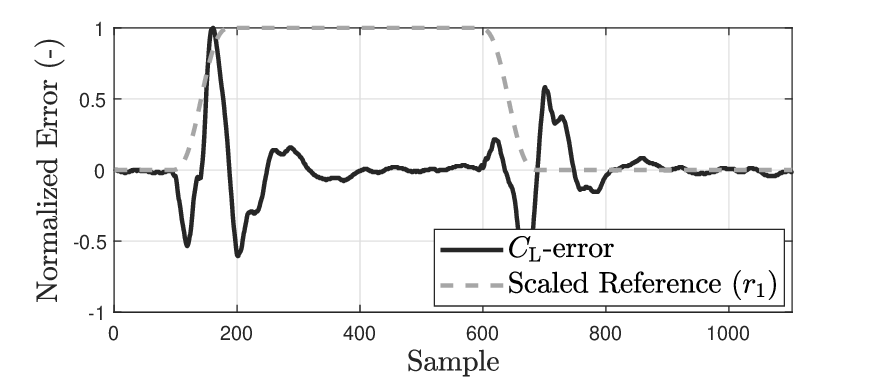}%
\label{fig: error linear 6mm}}
\hfil
\subfloat[]{\includegraphics[width=0.8\columnwidth]{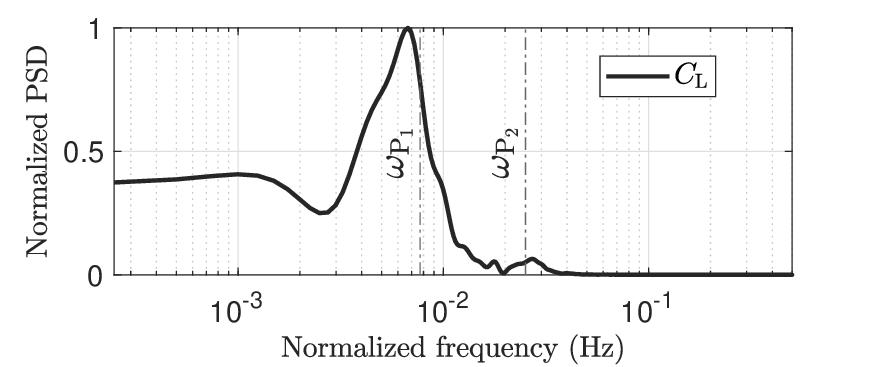}%
\label{fig: PSD linear 6mm}}
\caption{(a) Normalized error signal for both forward and backward motion under linear control. (b) Normalized PSD of the error within the stationary region.}
\label{fig: time-linear control 6mm}
\end{figure}

The designed LTI controller is implemented on the X-stage of the industrial wire bonder motion platform. Fig. \ref{fig: error linear 6mm} presents the error profile of the end-effector's position as it attempts to follow a trajectory $r_1(t)$, which resembles the typical trajectory depicted in Fig. \ref{Fig: Typical Reference}, with $\max\limits_{\forall t \in \mathbb{R}_{\geq0}}|r_1(t)|=r_{1,\mathrm{max}}$.
We define the stationary region as $[t_\text{r}, t_\text{e}]$ (see Fig. \ref{Fig: Typical Reference}), where $t_\text{r}$ represents the first time the reference signal reaches $r_\mathrm{max}$ (or 0 for backward motion), and $t_\text{e}$ denotes the last time it remains at that value (the moment just before beginning the next motion).

In Fig. \ref{fig: PSD linear 6mm}, the power spectral density (PSD) of the error is depicted, focusing on the samples within the stationary region $[t_\text{r}, t_\text{e}]$ after the forward motion is finished. The analysis reveals that the majority of the error energy is concentrated around $\omega_\mathrm{P_1}$, primarily corresponding to the frequency of the base frame vibration. This phenomenon arises from the reaction force exerted by the actuator on the base frame, which is subsequently transmitted to the end effector. This low-frequency vibration persists as a disturbance within the low-frequency range, remaining in the system even after the motion has concluded ($t \in [t_\text{r}, t_\text{e}]$). Notably, the peak of the PSD in Fig. \ref{fig: PSD linear 6mm} does not exactly coincide with \( \omega_{\mathrm{P_1}} \), as some error from the transient response remains and influences the location of the base frame vibration. By computing the PSD slightly after \( t_{\text{r}} \), these values would align more closely. However, in this study, we take the PSD from the moment the motion is first finished (\( t_{\text{r}} \)) for the sake of consistency throughout the analysis.

To suppress this base frame vibration, the magnitude of the sensitivity function at low frequencies must be reduced. However, due to the limitations of linear control—specifically, the waterbed effect—either the robust bounds would be violated, or the magnitude of the sensitivity at other frequencies would be amplified. Additionally, the analysis indicates an error arising from frequencies ($\omega_\mathrm{P_2}$) around the sensitivity peak (near the bandwidth), which is the region where the most energy transitions from the reference signal to the error ('$S = e/r$'). Consequently, this frequency range should also be considered as a potential source of excitation by different input references.

Given these constraints and the inherent limitations of linear control in further improving the performance of this motion platform, the subsequent section presents the design and implementation of a reset-based filter. This filter, informed by the results from Section \ref{Sec: CgLp}, aims to outperform the currently implemented linear controller.

\section{Reset Control Design for the wire bonder}\label{Sec: reset control of the wire bonder}
In this section, building upon the CgLp filter introduced in Section \ref{Sec: CgLp}, we aim to design a filter that can be integrated with the existing linear controller $C_\text{L}$. This filter is intended to reduce the magnitude of the sensitivity function around the frequencies of interest while ensuring that the magnitude of the function at other frequencies, particularly the robust bounds, remains unaffected.

In this study, we aim to incorporate an add-on nonlinear-based filter because it does not require any modifications to the existing linear control system. Moreover, the linear controller does not need to adapt to the designed nonlinear controller. This characteristic is particularly appealing for industrial applications where an LTI controller is already implemented, and the objective is to enhance its performance without compromising its existing functionality. In this context, we introduce the following performance improvement criterion, which provides a direct measure of the extent to which adding nonlinearity enhances system performance. Accordingly, we define the sensitivity improvement indicator ($\delta_{s}$) as:
\begin{equation}
    \label{eq: SII}
    \delta_{s}(\omega)=\frac{|S_\infty(\omega)|-|S(j\omega)|}{|S(j\omega)|}\%,
\end{equation}
where $S(j\omega)$ is the sensitivity of the linear control system and $S_\infty(\omega)$ is the (pseudo-) sensitivity (see \eqref{eq: S infty}) of the reset control system where a nonlinear controller has been added to the existing linear control system.

\subsection{Frequency Domain Filter Design}\label{subsec: frf filter design}
To reduce the magnitude of the sensitivity function around the base frame vibration frequencies ($\omega_\mathrm{P_1}$), we propose incorporating an inverse notch filter into the controller $C_\text{L}$. However, the inverse notch filter introduces negative phase shifts at frequencies beyond its effective range, which can reduce the phase margin and amplify the sensitivity function's peak magnitude. Both effects may lead to a violation of the constraints specified in \eqref{eq: constraints}.

To address this issue, we combine the inverse notch filter with the CgLp filter presented in Definition \ref{Lemma: New CgLp}. This combination compensates for the phase loss introduced by the inverse notch filter, thereby maintaining the phase margin and preserving the magnitude characteristics around the sensitivity peak. Thus, we design this add-on filter ($C_\text{g}$) as illustrated in Fig. \ref{Fig: Block diagram Add-on}. The $C_\text{N}$ represents the inverse notch filter, described as:
\begin{equation}
    \label{eq: inverse notch filter}
    C_\text{N}(s)=\frac{s^2/\omega_n^2+s/(\omega_n Q_1)+1}{s^2/\omega_n^2+s/(\omega_n Q_2)+1},
\end{equation}
where the peak of the inverse notch occurs at $\omega_n \in \mathbb{R}_{>0}$ with a magnitude of $\frac{Q_2}{Q_1} > 1$ ($Q_1, Q_2 \in \mathbb{R}_{>0}$).
\usetikzlibrary {arrows.meta}
\tikzstyle{sum} = [draw, fill=white, circle, minimum height=0.0em, minimum width=0.0em, anchor=center, inner sep=0pt]
\tikzstyle{block} = [draw,thick, fill=white, rectangle, minimum height=2.5em, minimum width=3em, anchor=center]
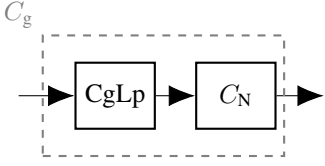
\begin{figure}[!t]
	\centering
	\begin{scaletikzpicturetowidth}{0.5\linewidth}
		\begin{tikzpicture}[scale=\tikzscale]
			\node[coordinate](input) at (0,0) {};
			
			\node[sum] (sum3) at (3.8,0) {};
			

			\node[block] (controller) at (2.7,0) {$C_\text{N}$};
			\node[block] (fo-higs) at (1.2,0) {CgLp};
			
			\node[coordinate](output) at (4.9,0) {};

			\draw[arrows = {-Latex[width=8pt, length=10pt]}] (input)  -- node[above]{} (fo-higs);
			\draw[arrows = {-Latex[width=8pt, length=10pt]}] (fo-higs) --node [above]{} (controller);
			\draw[arrows = {-Latex[width=8pt, length=10pt]}] (controller)  --node[above] {} (sum3);
            
            \draw [color=gray,thick,dashed](0.3,-0.75) rectangle (3.3,0.75);
			\node at (0.3,1) [left,color=gray] {$C_\text{g}$};
			
			
		\end{tikzpicture}
	\end{scaletikzpicturetowidth}
	\caption{The add-on filter ($C_\text{g}$) structure.}
	\label{Fig: Block diagram Add-on}	
\end{figure}

To specifically target the peak of the error energy at $\omega_\mathrm{P_1}$, we adopt the following procedure for designing the add-on filter $C_\text{g}$:
\begin{itemize} \item Select $\omega_n = \omega_\mathrm{P_1}$ with appropriate values for $Q_1$ and $Q_2$ to achieve sufficient width and height in the inverse notch filter's magnitude. (\(Q_1\) and \(Q_2\) should be chosen based on the magnitude and width of the PSD at the problematic frequency).
\item Determine the required phase of the CgLp filter as the phase lost due to the inverse notch filter at the bandwidth frequency (the open-loop crossover frequency, $\omega_\text{c}$) of the existing linear controller $C_\text{L}$:
\begin{equation} \label{eq: phase inverse notch} \theta_\text{CgLp}(\omega_\text{c}) = -\phase{C_\text{N}}(\omega_c). \end{equation}
\item Choose $\omega_l$ within the interval $[\omega_\mathrm{P_1}, \omega_\text{c}]$. This selection ensures that nonlinearity is not concentrated near the problematic frequency ($\omega_\mathrm{P_1}$), while still providing sufficient phase at the bandwidth frequency.
\item Select the desired value of $A_\rho$. (In this study, we consider $A_\rho = 0$ for all reset elements.) Use the required $\theta_\text{CgLp}$, along with the selected $\omega_l$ and $A_\rho$, to calculate the required $\omega_f$ (using Theorem \ref{lemma: wf calculation}).
\item Shape the CgLp components ($k_\text{c}$, $\mathcal{R}$, and $C_\mathfrak{c}$) based on the calculated $\omega_f$ (using Definition \ref{Lemma: New CgLp}).
\item Design the add-on filter as Fig. \ref{Fig: Block diagram Add-on}, based on the designed CgLp and $C_\text{N}$ filter.
\item Design the closed-loop control system as shown in Fig. \ref{Fig: Block diagram CL}, with \(C_1 = 1\), \(C_2 = k_\text{c} \cdot C_\mathfrak{c} \cdot C_\text{N} \cdot C_\text{L}\).
\item Compute the pseudo-sensitivity, \( S_\infty(\omega) \), for the designed reset-based controller and verify its compliance with the constraints presented in \eqref{eq: constraints}.
\item If both constraints in \eqref{eq: constraints} are satisfied and a sufficient reduction in \( |S_\infty(\omega)| \) at the problematic frequency is achieved, then the designed nonlinear controller is considered valid. However, if one or both constraints are violated, the frequency \(\omega_l\) may need to be reselected, or a less aggressive gain filter—specifically, the inverse notch filter (\(C_\text{N}\))—may be used instead. This adjustment is necessary because a less aggressive inverse notch filter introduces a smaller negative phase shift at the bandwidth frequency, thereby reducing the nonlinear action required from the CgLp element. As a result, the higher-order harmonics and \( |S_\infty(\omega)| \) around the bandwidth are mitigated, increasing the likelihood of satisfying the constraints and ensuring the validity of the controller.

\end{itemize}

Please note that the above steps can be applied to any other filter in place of $C_\text{N}$ by simply selecting a different filter in \eqref{eq: inverse notch filter}. Additionally, regarding Assumption \ref{Assumption: two reset per cycle}, and to obtain a more reliable approximation of the predicted closed-loop performance, we verify that the designed reset-based controller results in exactly two zero crossings per period for the signal \( e_r \) across the frequency range. Satisfying this additional condition does not necessarily confirm that Assumption \ref{Assumption: two reset per cycle} holds; however, it helps eliminate an obvious case of violation in this assumption and enhances the reliability of the results.

Considering the vibration frequency of the base frame depicted in Fig. \ref{fig: PSD linear 6mm} as the target frequency, we select the parameters for $C_\text{N}$ as $\omega_n=48.38\times10^{-3}\,$rad/sec ($\omega_n = \omega_\mathrm{P_1}$), $Q_1=1.31$, and $Q_2=1.62$. Having the bandwidth of the linear controller ($C_\mathrm{L}$) at $18.4\times10^{-2}\,$rad/sec, we take the aforementioned reset controller design steps, we select $A_\rho=0$, and $\omega_l=11.75\times10^{-2}\,$ rad/sec and calculate the rest of the parameters based on Definition \ref{Lemma: New CgLp} and Theorem \ref{lemma: wf calculation} and consequently the add-on filter.

It is also essential to highlight that, in this study, we are employing an optimal linear controller that is already operating at its limits, as can be observed in Fig. \ref{Fig: S linear}. Consequently, it becomes challenging to avoid violating the second constraint in \eqref{eq: constraints} after implementing the reset-based filter. To address this, we adopt a strategy that ensures compliance with the high-frequency constraint.

Specifically, in scenarios where the first constraint ($ M_\text{s} \leq 6 \, \text{dB}$) is satisfied, and a sufficient reduction is achieved around the desired frequency, but a minor violation occurs in $M_\text{r}$, we set the corner frequency of the lead element ($C_\mathfrak{c}$) as $\omega_f/c_\text{f}$, where $c_\text{f}\geq1$ is chosen to be close to one (e.g., $c_\text{f} \in [1,1.1]$). This results in a slight reduction in $|S_\infty(\omega)|$ around the resonance frequency ($\omega_\text{res}$), thereby ensuring compliance with the constraint on $M_\text{r}$.

It should be noted that this approach is effective only if the violation in $M_\text{r}$ is not significant. Otherwise, larger values of $c_\text{f}$ would be required, which could, in turn, impact $M_\text{s}$.

\begin{figure}[!t]
\centering
\includegraphics[width=0.8\columnwidth]{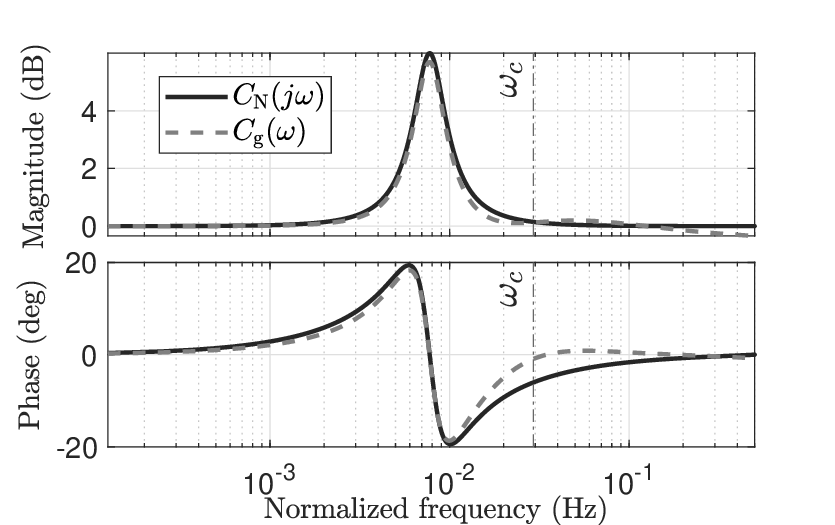}
\caption{The added gain to the system, with ($C_\text{g}$) and without ($C_\text{N}$) the CgLp part.}
\label{Fig: Anotch-CgLp}
\end{figure}

Fig. \ref{Fig: Anotch-CgLp} illustrates the inverse notch filter, \(C_\text{N}(j\omega)\), alongside the describing function of the add-on filter, \(C_\text{g}(\omega)\). The describing function of the add-on filter is defined as \(C_\text{g}(\omega) = \mathfrak{C}_1(\omega) \cdot C_\text{N}(j\omega)\). It can be observed that the filter \(C_\text{g}(\omega)\) exhibits nearly identical magnitude characteristics to \(C_\text{N}(j\omega)\) without introducing a negative phase at the bandwidth frequency. The effect of the proposed strategy (incorporation of $c_\text{f}$) is also evident in Fig. \ref{Fig: Anotch-CgLp}, where the gain of the add-on filter slightly decreases at high frequencies (with $c_\text{f}=1.06$).

The designed add-on filter is subsequently implemented to enhance the performance of the linear controller \(C_\text{L}\), thereby shaping the reset-based controller \({C_\text{NL}}_1\), with \(C_1 = 1\), \(C_2 = k_\text{c} \cdot C_\mathfrak{c} \cdot C_\text{N} \cdot C_\text{L}\), and \(\mathcal{R}\) functions as a proportional GFORE element defined by the parameters \((\mathcal{R}: A_r = -\omega_r, \, B_r = 1, \, C_r = \omega_r, \, D_r = \frac{\omega_l}{\omega_f - \omega_l})\).

\begin{figure}[!t]
\centering
\subfloat[]{\includegraphics[width=0.8\columnwidth]{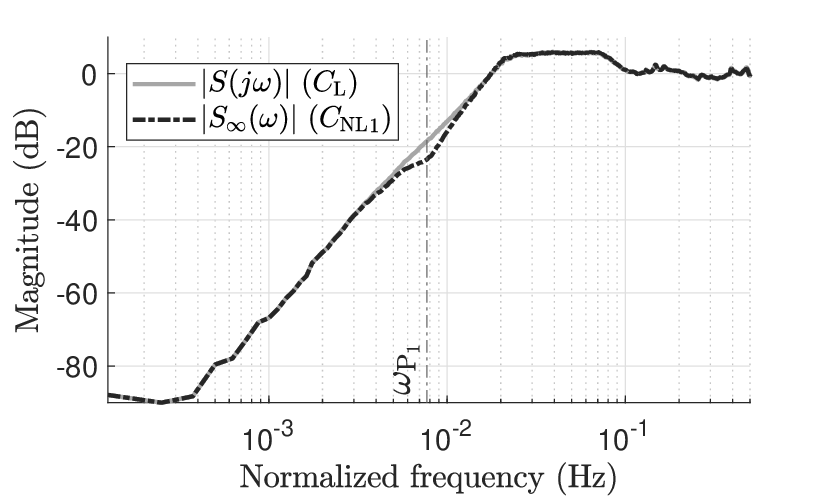}%
\label{fig: S_infty}}
\hfil
\subfloat[]{\includegraphics[width=0.8\columnwidth]{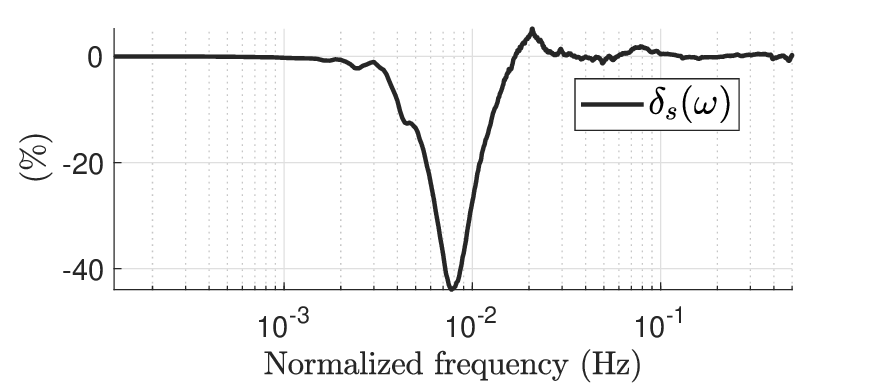}}%
\label{fig: Sii}
\caption{(a) The (pseudo-) sensitivity magnitude for ${C_\text{L}}$ and ${C_\text{NL}}_1$. (b) The Sensitivity Improvement Indicator for controller ${C_\text{NL}}_1$.}
\label{fig: S_infty and Sii}
\end{figure}

Utilizing Theorem \ref{Theorem: Sn theorem} and \eqref{eq: S infty}, we calculate the pseudo-sensitivity for the designed ${C_\text{NL}}_1$ controller.
In Fig. \ref{fig: S_infty}, the pseudo-sensitivity ($S_\infty (\omega)$) of the designed reset-based controller (${C_\text{NL}}_1$) is plotted alongside the sensitivity of the linear controller ($C_\text{L}$). It is observed that the magnitude of the sensitivity is reduced at the problematic frequency ($\omega_\mathrm{P_1}$) for the ${C_\text{NL}}_1$ controller, while all robustness constraints are satisfied for this controller. To better observe this, the Bode sensitivity integral has been calculated for both controllers as
\begin{equation}
    \int _{f_1}^{f_\text{end} }\ln |S(j\omega )|\text{d}\omega=9.83,
    \end{equation}
and
\begin{equation}
    \int _{f_1}^{f_\text{end} }\ln |S_\infty(\omega )|\text{d}\omega=-7.46,
\end{equation}
where $[f_1,f_\text{end}]$ is the frequency range where the plant is identified. The calculated Bode sensitivity integrals show that the reset-based controller (${C_\text{NL}}_1$) results in a reduction of $17.3$ compared to the linear controller.

To better observe the effect of the added filter on the sensitivity of its linear counterpart, we utilize the sensitivity improvement indicator ($\delta_{s}(\omega)$) presented in \eqref{eq: SII}. The $\delta_{s}(\omega)$ value, calculated using the sensitivities of ${C_\text{L}}$ and ${C_\text{NL}}_1$, is shown in Fig. \ref{fig: S_infty and Sii}b. This metric provides a clearer understanding of the impact of incorporating the nonlinear element in the closed-loop control system.

As depicted in Fig. \ref{fig: S_infty and Sii}b, negative values of $\delta_{s}(\omega)$ indicate a reduction in the energy of the error at corresponding frequencies, suggesting improved performance. It can be observed that, around the base-frame vibration frequency, the reset-based controller achieves approximately a 40 percent reduction in sensitivity magnitude compared to the linear controller. Conversely, positive values of $\delta_{s}(\omega)$ suggest that if those frequencies are excited, the nonlinearity introduced by the reset element could lead to higher error levels compared to the linear controller. This insight enables the designer to shape the controller more effectively by analyzing this FRF-based factor, thus avoiding error magnification around critical frequencies during the design process. 

After designing and validating the reset-based controller in the frequency domain, we aim to implement it on the physical wire bonder. Before proceeding, we outline practical guidelines for the discrete-time implementation of this reset-based controller. Finally, we validate its effectiveness through time-domain experiments.

\subsection{Practical Guidelines for the Implementation of Linear and Reset Controllers}
All linear and reset controllers discussed in this study are implemented in a digital framework. The discretization of all LTI elements is performed using the Tustin approximation method, which provides adequate phase preservation of the continuous-time system compared to other approximation techniques, especially up to frequencies close to the Nyquist frequency \cite{aastrom2013computerTustin}. Additionally, all frequency responses associated with linear elements are expressed in discrete-time (discretized FRF).
\usetikzlibrary {arrows.meta}
\tikzstyle{sum} = [draw, fill=white, circle, minimum height=0.6em, minimum width=0.6em, anchor=center, inner sep=0pt]
\tikzstyle{block} = [draw,thick, fill=white, rectangle, minimum height=2.5em, minimum width=3em, anchor=center]
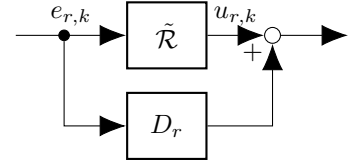
\begin{figure}[!t]
	\centering
	\begin{scaletikzpicturetowidth}{0.5\linewidth}
		\begin{tikzpicture}[scale=\tikzscale]
			\node[coordinate](input) at (0,0) {};

             \node[sum, fill=black, minimum size=0.4em] (dot2) at (0.8,0) {};
             
            \node[block] (R) at (2.5,0) {$\tilde{\mathcal{R}}$};

            \node[block] (Dr) at (2.5,-1.5) {$D_r$};
            
		\node[sum] (sumDr) at (4.25,0) {};

            \node[coordinate] (output) at (5.5,0) {};
			

			
			\draw[arrows = {-Latex[width=8pt, length=10pt]}] (input)  -- node[above]{$e_{r,k}$} (R);

            \draw[arrows = {-Latex[width=8pt, length=10pt]}] (dot2)  |- node[above]{} (Dr);
            
            \draw[arrows = {-Latex[width=8pt, length=10pt]}] (Dr)  -| node[pos=0.95,left]{$+$} (sumDr);

			\draw[arrows = {-Latex[width=8pt, length=10pt]}] (R)  --node[above] {$u_{r,k}$} (sumDr);

            \draw[arrows = {-Latex[width=8pt, length=10pt]}] (sumDr)  --node[above] {} (output);
			
			
		\end{tikzpicture}
	\end{scaletikzpicturetowidth}
	\caption{The discrete-time implementation of the proportional GFORE element.}
	\label{Fig: Block diagram discrete}	
\end{figure}

The discrete-time realization of the reset element in \eqref{eq: reset state space} (for the first-order reset element, $n_r=1$) is presented as Fig. \ref{Fig: Block diagram discrete}. Since $D_r$ in \eqref{eq: reset state space} represents only the feedthrough term, which can be placed in parallel with the GFORE element, we provide the discrete-time realization for the case where $D_r=0$ and subsequently add $D_r$ to the discretized output. Thus, we obtain:

\begin{equation}
\squeezespaces{0.5}
\tilde{\mathcal{R}} := 
\begin{cases}
    x_{r,k+1} = \tilde{A_r}x_{r,k} + \tilde{B_r}e_{r,k}, & \text{if } (e_{r,k}, e_{r,k-1}) \notin \tilde{\mathcal{F}} \\
   x_{r,k+1} =  A_\rho\left(\tilde{A_r} x_{r,k} + \tilde{B_r}e_{r,k}\right), & \text{if } (e_{r,k}, e_{r,k-1}) \in \tilde{\mathcal{F}} \\
    u_{r,k} = \tilde{C}_rx_{r,k}+\tilde{D}_re_{r,k}, & \text{if } (e_{r,k}, e_{r,k-1}) \notin \tilde{\mathcal{F}}\\
    u_{r,k} = A_\rho\left(\tilde{C}_r  x_{r,k}+\tilde{D}_r e_{r,k}\right), & \text{if } (e_{r,k}, e_{r,k-1}) \in \tilde{\mathcal{F}}\\
\end{cases}
\label{eq: GFORE discrete}
\end{equation}
with
\begin{equation}
\label{eq: discrete M}
 \tilde{\mathcal{F}} := \left\{ (e_{r,k}, e_{r,k-1})\in \mathbb{R}^2 \mid e_{r,k} = 0 \, \vee \, e_{r,k} e_{r,k-1} < 0 \right\},  
\end{equation}
where $\tilde{A_r}$, $\tilde{B_r}$, $\tilde{C_r}$, and $\tilde{D_r}$ all $\in \mathbb{R}$, represent the state-space matrices of the Tustin-discretized base linear system (GFORE with $A_r$, $B_r$, $C_r$, and $D_r=0$), and $k \in \mathbb{N}$ denotes the sample index. The system primarily follows the Tustin-discretized linear dynamics. However, if a reset is detected, the state and the output are reset to $A_\rho$ times their original values.

It should be noted that in this study, we used a different and more precise reset surface compared to other studies, such as \cite{caporale2024practical}. Specifically, the system in \eqref{eq: GFORE discrete} resets its state \( x_{r,k+1} \) when either \( e_{r,k} = 0 \) or \( e_{r,k} e_{r,k-1} < 0 \). 
A limitation of the previous discrete reset surface, as defined by the condition 
\[
\tilde{\mathcal{F}} := \left\{ (e_{r,k}, e_{r,k-1})\in \mathbb{R}^2 \mid e_{r,k} e_{r,k-1} \leq 0 \right\},
\]  
presented in \cite{caporale2024practical}, was that it triggered a reset at both sample \( k-1 \) and \( k \) when \( e_{r,k-1} = 0 \). This behavior resulted in two resets occurring for a single zero crossing.

It is important to emphasize that, although the actual zero-crossing occurs between samples \( k-1 \) and \( k \), approximating sample \( k \) as the reset event sample and performing discretization do not introduce significant inaccuracies in continuous-time HOSIDF analysis (the method used in this work to represent the reset element in the frequency domain). This is because, in this study, the reset element is considered as a proportional GFORE with a nonzero feedthrough term. As a result, the nonlinearity is effectively attenuated well below the sampling frequency (\(\omega_f \approx \frac{F_s}{30}\), where \( F_s \) denotes the sampling frequency). Furthermore, since the bandwidth frequency is located approximately around \( \omega_f \), it is reasonable to assume that this approximation does not introduce significant errors, particularly in terms of DF-based phase analysis around the bandwidth frequency.

With the reset element digitally implemented, we next present the experimental results of applying reset-based controllers to the wire bonder.

\section{Experimental Results}\label{sec: experimental result}
\begin{figure}[!t]
\centering
\subfloat[]{\includegraphics[width=0.8\columnwidth]{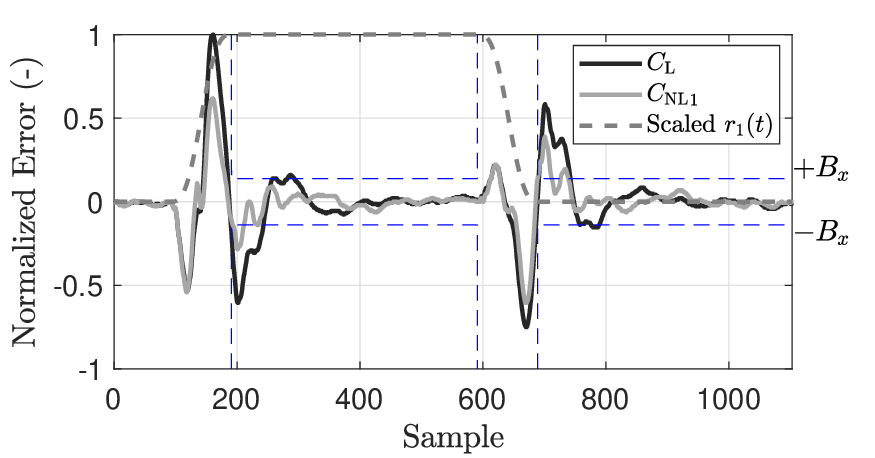}%
\label{fig: Error NL 6mm}}
\hfil
\subfloat[]{\includegraphics[width=0.8\columnwidth]{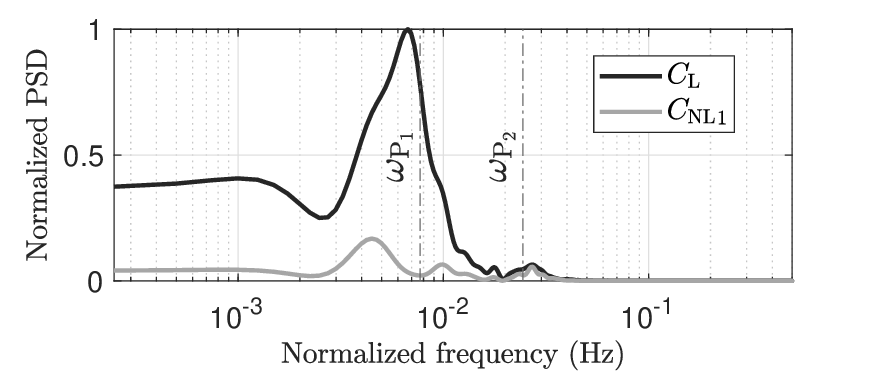}}%
\label{fig: PSD NL 6mm}
\caption{(a) Normalized error signal for both forward and backward motion for linear and nonlinear controllers. (b) Normalized PSD of the error within the stationary region of forward motion.}
\label{fig: Error and PSD NL 6mm}
\end{figure}

In this section, we implement the designed reset controller, ${C_\text{NL}}_1$, on the physical wire bonder, considering the reference trajectory $r_1(t)$. A comparison is then made with its linear counterpart controller. Fig. \ref{fig: Error NL 6mm} presents the error profiles for both ${C_\text{L}}$ and ${C_\text{NL}}_1$. A clear reduction in the error is observed. 
To quantify this performance, we define \( T^\star \in \mathbb{R}_{>0} \) as the time interval from the moment the motion is completed (\( t_\text{r} \), as depicted in Fig. \ref{Fig: Typical Reference}) until the error last remains outside the \( \pm B_\mathrm{x} \) bound. Evidently, the objective is to minimize this region, enabling the system to transition more quickly to the bonding phase, thereby increasing the overall bonding efficiency. Mathematically, $T^\star$ is expressed as:
\begin{equation}
    \label{eq: Ts}
    T^\star=t_\text{s}-t_\text{r},
\end{equation}
where $t_\text{s}\in \mathbb{R}_{>0}$ represents the settling time, defined as:
\begin{equation}
    \label{eq: ts}
    t_\text{s}=\min\limits_{}\{t\in [t_\text{r},t_\text{e}] \Big|\, |e(t)|\leq B_\mathrm{x} \, \forall \, t\geq t_\text{s} \}.
\end{equation}

In Fig. \ref{fig: Error NL 6mm}, the $\pm B_\mathrm{x}$ bound is scaled relative to the error. By calculating $T^\star$ for both the linear and nonlinear cases, it is demonstrated that the reset-based filter ${C_\text{NL}}_1$ reduces the settling duration ($T^\star$) by $80.4\%$ for forward motion and $50.5\%$ for backward motion compared to the linear controller ${C_\text{L}}$.

In the linear control setup, when the end-effector reached its final position, vibrations in the base frame necessitated waiting for the oscillations to dampen before the system could stabilize within the defined boundary. This delay was critical to ensure precise connections and avoid collisions with adjacent wires. By incorporating the reset-based filter into the control loop without altering any mechanical components of the machine, the system achieved significantly faster transitions to the bonding phase by effectively damping unwanted vibrations. Notably, since these forward and backward motions occur multiple times per second, the $80.4\%$ improvement in forward motion and $50.5\%$ improvement in backward motion significantly enhance the efficiency of the machine. Consequently, this leads to a substantial increase in the number of microchips packaged.

The PSD of the errors is also calculated for the stationary region $[t_\text{r}, t_\text{e}]$ and is depicted in Fig. \ref{fig: Error and PSD NL 6mm}b. For simplicity, only the PSD corresponding to the forward motion is plotted. It is evident that the energy of the error is significantly reduced at the base-frame vibration frequency ($\omega_\mathrm{P_1}$). This observation demonstrates how the waterbed effect has been mitigated, enabling the reduction of error at a specific frequency without compromising performance at other frequencies. To quantify this performance, the root mean square (RMS) of the error is calculated for the stationary region $[t_\text{r}, t_\text{e}]$. The results indicate a 58.9\% reduction in RMS error for forward motion and a 46.4\% reduction for backward motion. The reduction in the RMS value is as important as the reduction in the settling period. While the goal of minimizing the settling period was to enable the system to reach the bonding region as quickly as possible, the reduction in RMS error demonstrates that the end-effector remains in the target position with a significantly lower deviation from the desired point. This improvement contributes to higher accuracy during the bonding process. Consequently, the incorporation of the reset-based filter resulted in enhancements to both the speed and accuracy of this wire bonder machine.

Thus far, we have analyzed the control challenges associated with trajectory tracking of the reference signal \( r_1(t) \). The problematic frequency (\( \omega_{\mathrm{P_1}} \)) identified may arise exclusively for this specific reference. Therefore, in the subsequent section, we present a robust reset controller capable of enhancing system performance for tracking various potential reference trajectories.\\

\subsubsection*{Robust control design}\label{subsec: robust control design}
In Section \ref{subsec: linear control for wire bonder}, we illustrated and discussed the presence of base-frame vibrations in the error signal while the system attempted to follow the reference trajectory \( r_1(t) \). During the wire bonding process, the end-effector typically moves from one point to another. These movements can be categorized as short-range, mid-range, or long-range motions. Although \( r_1(t) \) is scaled in Section \ref{subsec: linear control for wire bonder} for confidentiality reasons, it represents a long-range reference trajectory. The energy spectrum of these long-range motion setpoints primarily consists of low-frequency components, which excite base-frame vibrations and subsequently affect the position of the end-effector, leading to error concentration at \( \omega_{\mathrm{P_1}} \).

\begin{figure}[!t]
\centering
\subfloat[]{\includegraphics[width=0.8\columnwidth]{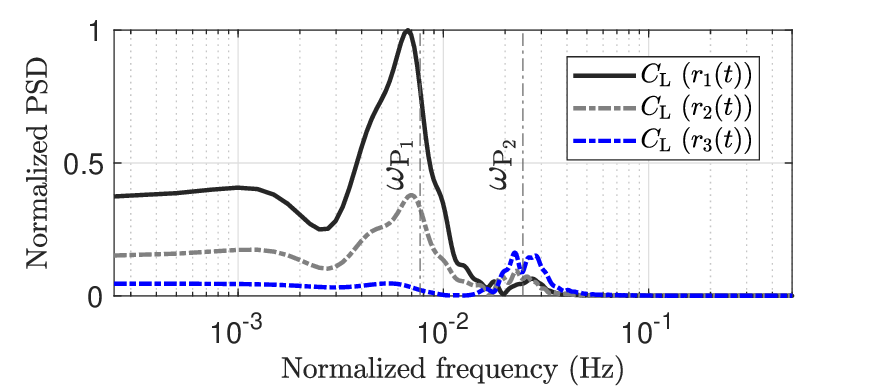}%
\label{fig: PSD different reference}}
\hfil
\subfloat[]{\includegraphics[width=0.8\columnwidth]{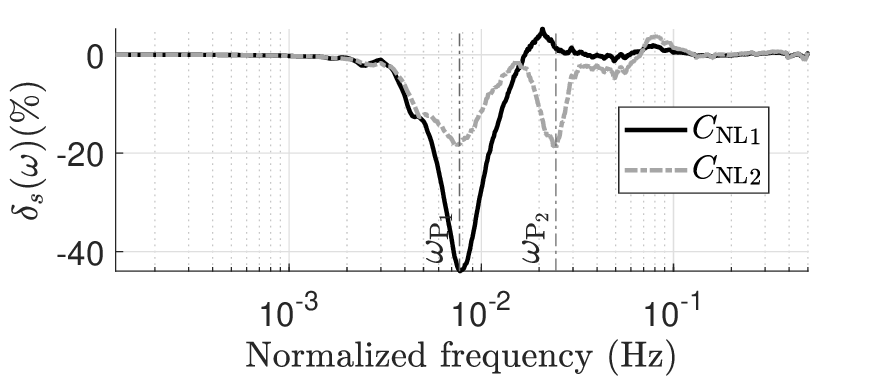}}%
\label{fig: SII skewed}
\caption{(a) The PSD of the error for the controller ${C_\text{L}}$ following $r_1(t)$, $r_2(t)$, and $r_3(t)$. (b) The sensitivity improvement indicator for the controller \( {C_\text{NL}}_2 \), designed to target multiple problematic frequencies.}
\label{fig: PSD and Sii skewed}
\end{figure}

However, as the motion range decreases, the energy associated with the error gradually shifts from base-frame vibrations (\( \omega_{\mathrm{P_1}} \)) to vibrations near the bandwidth frequency (\( \omega_{\mathrm{P_2}} \)). This shift occurs because the bandwidth frequency exhibits the highest energy levels, as indicated by the sensitivity analysis. Additionally, short-range input reference signals typically contain high-frequency components in their energy spectrum, which have a greater impact on error at higher frequencies during short movements. To further investigate this behavior, we study two additional reference trajectories, \( r_2(t) \) and \( r_3(t) \), where
\[
\max\limits_{\forall t \in \mathbb{R}_{\geq0}} |r_2(t)| = r_{2,\mathrm{max}} = r_{1,\mathrm{max}}/2,
\]
and
\[
\max\limits_{\forall t \in \mathbb{R}_{\geq0}} |r_3(t)| = r_{3,\mathrm{max}} = r_{1,\mathrm{max}}/10.
\]
It should be noted that the reference trajectories \( r_2(t) \) and \( r_3(t) \) are not only shorter in distance but also shorter in duration compared to \( r_1(t) \). The linear controller \( C_\text{L} \) is implemented for both reference trajectories \( r_2(t) \) and \( r_3(t) \). The PSD of the error regarding the three different reference trajectories is illustrated in Fig. \ref{fig: PSD different reference}. It is observed that the magnitude of the PSD for \( r_2(t) \) is distributed across both \( \omega_{\mathrm{P_1}} \) and \( \omega_{\mathrm{P_2}} \) frequencies, whereas for \( r_3(t) \), the magnitude of the PSD is concentrated almost entirely at \( \omega_{\mathrm{P_2}} \). Consequently, the nonlinear controller \( {C_\mathrm{NL}}_1 \) may not perform optimally for reference trajectories \( r_2(t) \) and \( r_3(t) \), as it is designed to reduce error around the \( \omega_{\mathrm{P_1}} \) frequency and following \( r_2(t) \) and \( r_3(t) \) with this controller might not result in any improvement at \( \omega_{\mathrm{P_2}} \) frequency.

Thus, we consider a gain filter that combines two inverse notches (each has the same transfer function as \eqref{eq: inverse notch filter}), one at \( \omega_{\mathrm{P_1}}=48.38\times10^{-3}\, \)rad/sec and one at \( \omega_{\mathrm{P_2}}=15.33\times10^{-2}\, \)rad/sec. With this filter, denoted as \(C_\text{N}(s)=C_\mathrm{N_1}(s)C_\mathrm{N_2}(s)\), and by selecting $\omega_l=14.39\times10^{-2}\,$rad/sec, $A_\rho=0$, $Q_1=1.12$, $Q_2=1.59$ (for $C_\mathrm{N_1}$), and $Q_1=1.43$, $Q_2=1.59$ (for $C_\mathrm{N_2}$), we can follow the exact steps outlined in Section \ref{subsec: frf filter design} (equation \eqref{eq: inverse notch filter} and onward) to design the new nonlinear controller. Based on the design steps and considering the sequence of the elements as \(C_1 = C_\mathrm{N_2}\) (We had to move the \( C_\mathrm{N_2} \) filter before the reset element due to a firmware limitation on the maximum order of filters that can be implemented in $C_2$), \(C_2 = k_\text{c} \cdot C_\mathfrak{c} \cdot C_\mathrm{N_1} \cdot C_\text{L}\), a valid controller (\( {C_\text{NL}}_2 \)) is selected, which satisfies all constraints in \eqref{eq: constraints}, and results in the reduction of $|S_\infty(\omega)|$ in both \( \omega_{\mathrm{P_1}} \) and \( \omega_{\mathrm{P_2}} \).

The sensitivity improvement indicator for the controller \( {C_\text{NL}}_2 \) is depicted in Fig. \ref{fig: PSD and Sii skewed}b. It can be observed that it does not reduce the error at \( \omega_{\mathrm{P_1}} \) as effectively as the controller \( {C_\text{NL}}_1 \), but it demonstrates a reduction in sensitivity at \( \omega_{\mathrm{P_2}} \). This makes \( {C_\mathrm{NL}}_2 \) a better choice when the system must follow all trajectories in a single operation, ensuring that the error can be reduced compared to the linear controller, independent of the problematic frequency that is excited.

Following the design of \( {C_\text{NL}}_2 \), it is implemented on the physical wire bonder for various setpoints \( r_1(t) \), \( r_2(t) \), and \( r_3(t) \). Table \ref{Tabel: Experimental result all setpoints} presents the results for the two designed nonlinear controllers (\( {C_\text{NL}}_1 \) and \( {C_\text{NL}}_2 \)), highlighting their performance in terms of settling time and RMS error. These metrics are compared to those of the linear controller, with the improvements expressed as percentages (minus values mean the improvement compared to the linear controller).

It can be observed that while the controller \( {C_\text{NL}}_1 \) significantly reduces both the settling time and the RMS error for the reference trajectory \( r_1(t) \), it does not outperform the linear controller \( {C_\text{L}} \) for other range of motions. Therefore, in scenarios involving a combination of both short- and long-range motions, or where a controller needs to perform independently of the reference trajectory while still improving system performance, the controller \( {C_\text{NL}}_2 \) should be selected. As shown in Table \ref{Tabel: Experimental result all setpoints}, this controller consistently outperforms the linear controller across all reference trajectories in terms of both settling duration and RMS error reduction. 

It is noteworthy that these improvements were achieved solely through the frequency-domain tuning method, leveraging pseudo-sensitivity and the newly introduced sensitivity improvement factor. This approach allows designers to shape $\delta_{s}(\omega)$ based on the power spectrum density analysis of a pre-designed linear controller and enhance its performance without altering the controller itself. This method is particularly advantageous in industrial applications, as it enables the frequency-domain designing and shaping of a nonlinear-based filter to improve system performance, even without access to the parametric model of the system or detailed parameters of the implemented linear controller.

\begin{table}[!t]
\caption{Improvement values for settling period and RMS error for \( {C_\text{NL}}_1 \) and \( {C_\text{NL}}_2 \) compared to \( {C_\text{L}} \).}\label{Tabel: Experimental result all setpoints}
\resizebox{\columnwidth}{!}{%
    \begin{tabular}{|c|c|c|c|c|}
    \hline
    &\multicolumn{2}{c|}{\( {C_\text{NL}}_1 \)}& \multicolumn{2}{c|}{\( {C_\text{NL}}_2 \)}\\
    \hline
    Reference & \makecell{$T^\star$ \\ Change (\%)}& \makecell{RMS error \\ Change (\%)}& \makecell{$T^\star$ \\ Change (\%)}& \makecell{RMS error \\ Change (\%)} \\
    \hline
    \makecell{$r_1(t)$-Forward \\ $r_1(t)$-Backward}&\makecell{-80.4\\-50.5}&\makecell{-58.9\\-46.4}&\makecell{-54.9\\-51.5}&\makecell{-51.4\\-39.9}\\ \hline
    \makecell{$r_2(t)$-Forward \\ $r_2(t)$-Backward} &\makecell{+2.9\\+0.7}&\makecell{-31.6\\-19.3}&\makecell{-25.7\\-28.8}&\makecell{-32.6\\-21.9}\\ \hline
    \makecell{$r_3(t)$-Forward \\ $r_3(t)$-Backward}& \makecell{-0.8\\-0.3}&\makecell{+5.3\\-1.8}&\makecell{-5.1\\-3.4}&\makecell{-23.7\\-21.9}\\ \hline
    Average&-21.4&-25.5&-28.2&-31.9 \\ \hline
    \end{tabular}%
    }
 \end{table}

\section{Conclusion}\label{Sec: Conclusion}
In this article, we demonstrated the design of a reset-based controller and its integration into an existing control loop without modifying other constraints or compromising the system’s performance criteria. This was achieved through the introduction of a proportional first-order reset element and the formulation of the CgLp filter based on this element. By employing this first-order reset element with a non-zero feedthrough term, the nonlinear filter benefits could be exploited while mitigating the effects of nonlinearity beyond a certain frequency. The proposed formulation and combination of the CgLp element with any required gain-based filter (e.g., CgLp + inverse notch) allow for direct integration into an existing linear controller. This feature is particularly valuable in industrial applications where redesigning the entire control loop is often impractical.

Shaping this nonlinear filter relies on a closed-loop frequency-domain method, allowing designers to tailor the controller using only the plant’s frequency response function. The sensitivity improvement indicator was introduced to assess the impact of added nonlinearity directly in the closed-loop response. 

To validate this approach, we addressed a base-frame vibration problem in an industrial wire bonder. The objective was to resolve this issue without altering the mechanical structure or the pre-implemented linear controller while ensuring all robustness constraints were met. Experimental results demonstrated that the error energy was reduced to half of that achieved with the linear controller, and the bonding process speed was increased. To further generalize and enhance the robustness of the design, another reset-based filter was shaped using the sensitivity improvement indicator. This filter effectively targeted a broader frequency range, reducing error energy from very low frequencies up to the bandwidth frequency while maintaining robustness. Ultimately, the robust reset controller improved both the RMS error and the settling time across various reference trajectories.

In conclusion, this study introduces a step-by-step design method for a phase generator reset-based filter that mitigates the negative effects of nonlinearity. The resulting add-on filter can be used to shape the control loop gain constructively without altering the existing structure, making it compatible with any linear controller. 

For future work, the proposed design method could be extended and formalized as an optimization problem, where the cost functions minimize the sensitivity improvement indicator at problematic frequencies.
\appendices
\section{CgLp Construction}\label{Appendix: CgLp lemma}
To ensure that a GFORE element \(\left(A_r = -\omega_r, B_r = 1, C_r = \omega_r, D_r = 0\right)\) exhibits the same magnitude characteristics at low frequencies (\(\omega \to 0\)) and high frequencies (\(\omega \to \infty\)) as a linear low-pass filter \(\left(\frac{1}{1+{j\omega}/\omega_l}\right)\), similar to the approach in \cite[Section 3]{LukeIFAC2024}, the following conditions must hold:
\begin{equation}
\label{eq: H goes to 1}
    \left| H_1(\omega) \right|_{\omega \to 0} = 1,
\end{equation}
and
\begin{equation}
\label{eq: H1 omega infty}
    \left| H_1(\omega) \right|_{\omega \to \infty} = \frac{\omega_l}{\omega}.
\end{equation}
From Theorem \ref{theorem: reset hosidf}, the magnitude \(|H_1(\omega)|\) for the given GFORE parameters can be expressed as:
\begin{equation}
\label{eq: 36 app1}
|H_1(\omega)|=\frac{\omega_r\sqrt{1+\Theta_\text{D}^2(\omega)}}{\sqrt{\omega_r^2+\omega^2}}.
\end{equation}
Given that \(\Theta_\text{D}(0)=0\), \eqref{eq: 36 app1} yields  
\[
\left| H_1(\omega) \right|_{\omega \to 0} = 1,
\]  
which confirms that \eqref{eq: H goes to 1} holds. Furthermore, since  
\[
\lim_{\omega\rightarrow \infty}\Theta_\text{D}(\omega) = \frac{4(1-A_\rho)}{\pi(1+A_\rho)},
\]  
evaluating \eqref{eq: 36 app1} as \(\omega \to \infty\) and equating it to \eqref{eq: H1 omega infty} provides the expression for \(\omega_r\):  
\begin{equation}
\label{eq: corner frequency2}
\omega_r = \frac{\omega_l}{\sqrt{1 + \left(\frac{4(1 - A_\rho)}{\pi(1 + A_\rho)}\right)^2}}.
\end{equation}

In this proof, we demonstrate that employing the CgLp configuration presented in Definition \ref{Lemma: New CgLp} leads to the following result:
\begin{itemize}
    \item \( \left| \mathfrak{C}_1(\omega) \right|_{\omega \to 0} = 1 \),
    \item \( \left| \mathfrak{C}_1(\omega) \right|_{\omega \to \infty} = 1 \).
\end{itemize}
Thus, from \eqref{eq: CgLP HOSIDF}, we have:
\begin{equation}\label{eq: w goes 0}
\left|\mathfrak{C}_1(\omega)\right|= k_\text{c}\left|H_1(\omega)\right|\left|C_\mathfrak{c}(j\omega)\right|.
\end{equation}
Using \( H_1(\omega) \) from \eqref{eq reset HOSIDF}, this time for \( A_r = -\omega_r \), \( B_r = 1 \), \( C_r = \omega_r \), and \( D_r\neq0 \), we get:
\begin{equation}
\left|H_1(\omega)\right|=\left|\omega_r\left(j\omega + \omega_r\right)^{-1}\left(1 + j\Theta_\text{D}(\omega)\right)+D_r\right|,
\end{equation}
which results in:
\begin{equation}\label{eq: proof 0}
    \lim_{\omega \rightarrow 0} k_\text{c}\left|H_1(\omega)\right|\left|C_\mathfrak{c}(j\omega)\right|=k_\text{c}(D_r+1),
\end{equation}
and using \( k_\text{c} \) from \eqref{eq: kc} and \( D_r \) from \eqref{eq: Dr}, we obtain:
\begin{equation}
    \left| \mathfrak{C}_1(\omega) \right|_{\omega \to 0} = 1.
\end{equation}
To analyze \( \left| \mathfrak{C}_1(\omega) \right|_{\omega \to \infty} \), we have:
\begin{equation}\label{eq: w to infty}
    \lim_{\omega \rightarrow \infty} k_\text{c}\left|H_1(\omega)\right|\left|C_\mathfrak{c}(j\omega)\right|=k_\text{c}D_r\frac{\omega_f}{\omega_l},
\end{equation}
where, by substituting \( k_\text{c} \) from \eqref{eq: kc} and \( D_r \) from \eqref{eq: Dr}, we obtain:
\begin{equation}\label{eq: proof infty}
    \left| \mathfrak{C}_1(\omega) \right|_{\omega \to \infty} = 1.
\end{equation}
From \eqref{eq: proof 0} and \eqref{eq: w to infty}, it can also be observed that the values of \( k_\text{c} \) and \( D_r \) provided in Definition \ref{Lemma: New CgLp} are the only possible values that satisfy both \( k_\text{c}(D_r+1)=1 \) and \( k_\text{c}D_r\frac{\omega_f}{\omega_l}=1 \). \qed

\section{Proof of Theorem \ref{lemma: wf calculation}} \label{Appendix: wf calculation} Given the phase of the describing function of the \( \text{CgLp} \) as
\begin{equation}
\label{eq: theta cglp2}
\begin{split}
\theta&_\mathrm{CgLp}(\omega) = \\
&\arctan{\left(\frac{b}{a+\frac{\omega_l}{\omega_f-\omega_l}}\right)} + \arctan{\left(\frac{\omega}{\omega_l}\right)} - \arctan{\left(\frac{\omega}{\omega_f}\right)},
\end{split}
\end{equation}
we can express
\begin{equation}
\begin{split}
\theta_\mathrm{CgLp}(\omega) - \arctan&{\left(\frac{\omega}{\omega_l}\right)} = \\
&\arctan{\left(\frac{b}{a+\frac{\omega_l}{\omega_f-\omega_l}}\right)} - \arctan{\left(\frac{\omega}{\omega_f}\right)}.
\end{split}
\end{equation}
Using the identity \(\arctan{(x)} - \arctan{(y)} = \arctan{\left(\frac{x-y}{1+xy}\right)}\), we obtain:
\begin{equation}
\label{eq: sum arctan}
\begin{split}
\theta_\mathrm{CgLp}(\omega) -& \arctan{\left(\frac{\omega}{\omega_l}\right)} = \\
&\arctan{\left(\frac{b\omega_f^{2} - \omega_f(b\omega_l + a\omega) + (a-1)\omega\omega_l}{a\omega_f^{2} - \omega_f((a-1)\omega_l - b\omega) - b\omega\omega_l}\right)}.
\end{split}
\end{equation}
We define
\begin{equation}
    Q=\tan{\left(\theta_\mathrm{CgLp}(\omega)-\arctan{\left(\frac{\omega}{\omega_l}\right)}\right)},
\end{equation}
which from \eqref{eq: sum arctan} can be written as
\begin{equation}
\label{eq: Q wf2}
     Q={\left(\frac{b\omega_f^{2}-\omega_f(b\omega_l+a\omega)+(a-1)\omega\omega_l}{a\omega_f^{2}-\omega_f((a-1)\omega_l-b\omega)-b\omega\omega_l}\right)}.
\end{equation}
Expanding \eqref{eq: Q wf2} results in the following second-order equation:
\begin{equation}
    \label{eq: expanded wf2}
    \begin{split}
        &\left(aQ-b\right)\omega_f^2+\left(b\omega Q+b\omega_l+a\omega-(a-1)\omega_lQ\right)\omega_f\\
        &-\omega\omega_l(bQ+a-1)=0.
    \end{split}
\end{equation}
Solving the equality in \eqref{eq: expanded wf2} gives two values for $\omega_f$ as
\begin{equation}
\label{eq: omega_f in proof}
    \omega_{f_1}=\frac{-k_2-\sqrt{k_2^2-4k_1k_3}}{2k_1},
\end{equation}
and
\begin{equation}
\label{eq: omega_f in proof2}
    \omega_{f_2}=\frac{-k_2+\sqrt{k_2^2-4k_1k_3}}{2k_1},
\end{equation}
where
\begin{equation}
    \begin{split}
        k_1&=aQ-b, \\
        k_2&=b\omega Q+b\omega_l+a\omega-(a-1)\omega_lQ, \\
        k_3&=-\omega\omega_l[bQ+a-1].\\
    \end{split}
\end{equation}
Since we consider the required phase as $\theta_\text{CgLp}(\omega) \in (0, \theta_\text{M}(\omega))$, it follows from Lemma \ref{lemma: max theta} that there is at least one solution for $\omega_f \in [\omega_l, \infty)$ that satisfies the given phase requirement. The second solution for $\omega_f$, may lie either within or outside the interval $[\omega_l, \infty)$.  

If two solutions exist within $[\omega_l, \infty)$, the smaller one is chosen, as terminating the nonlinear integrator action earlier reduces the impact of HOSIDFs. Conversely, if the second solution falls outside the interval $[\omega_l, \infty)$, ($\omega_f < \omega_l$), the other solution ($\omega_f \geq \omega_l$) is selected. This results in:
\begin{equation}
\omega_f=
\begin{cases}
\min(\omega_{f_1},\omega_{f_2}), & \text{if both} \,\omega_{f_1},\omega_{f_2} \in [\omega_l,\infty),\\
\max(\omega_{f_1},\omega_{f_2}), & \text{otherwise}.
\end{cases}
\end{equation}
\qed

\bibliography{ref}
\bibliographystyle{IEEEtran}

\vspace{11pt}

\begin{IEEEbiography}[{\includegraphics[width=1in,height=1.25in,clip,keepaspectratio]{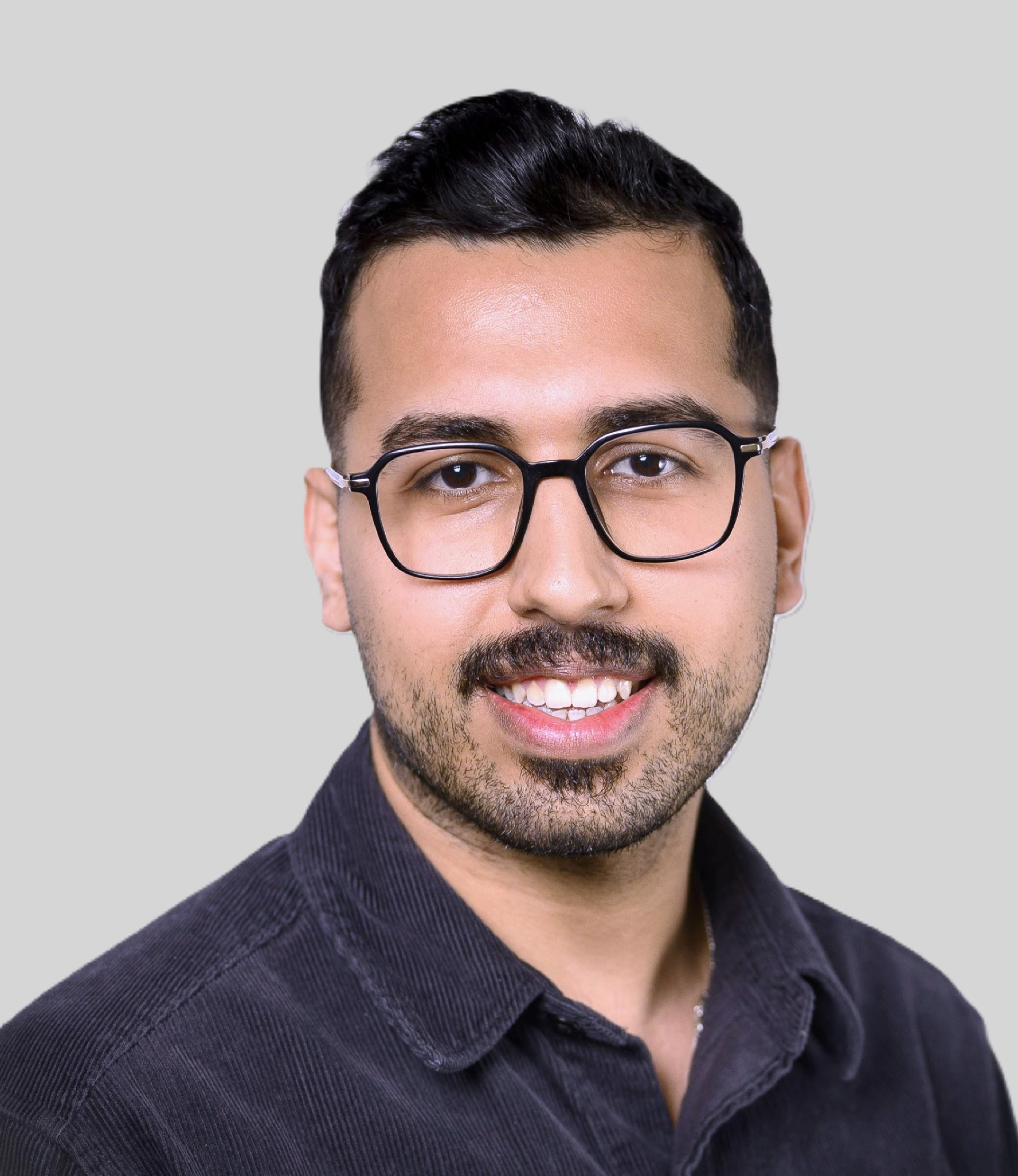}}]{S. Ali Hosseini}
received his M.Sc. degree in Systems and Control Engineering, specializing in nonlinear control (with a focus on hybrid integrator-gain systems), from Sharif University of Technology, Tehran, Iran, in 2022.

He is currently pursuing a Ph.D. in the Department of Precision and Microsystems Engineering at Delft University of Technology, Delft, The Netherlands. His research focuses on addressing industrial control challenges using nonlinear control techniques in close collaboration with ASMPT, Beuningen, The Netherlands. His research interests include precision motion control, nonlinear control systems (such as reset and hybrid systems), and mechatronic system design.

\end{IEEEbiography}

\vspace{11pt}

\begin{IEEEbiography}[{\includegraphics[width=1in,height=1.25in,clip,keepaspectratio]{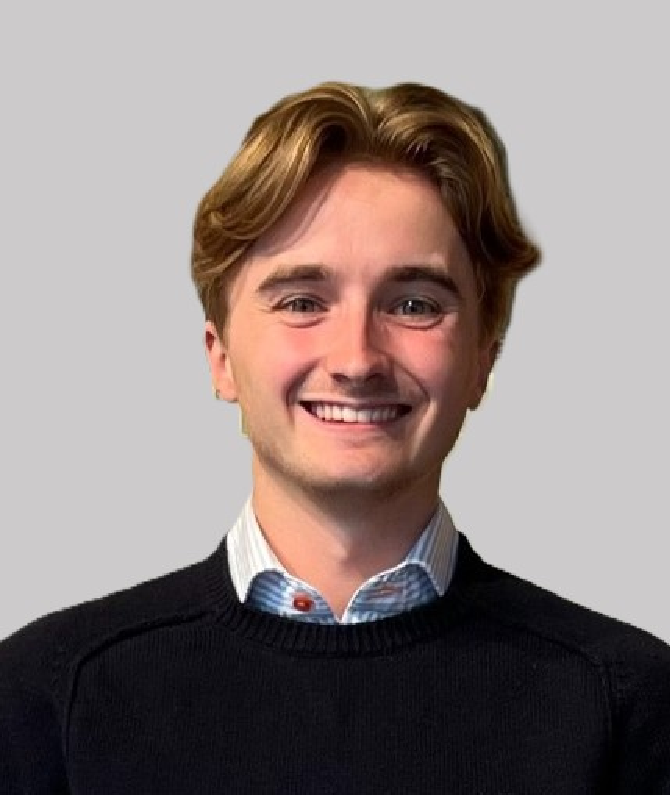}}]{Fabian R. Quinten} received the M.Sc. degree in mechanical engineering with specialization in mechatronics from Delft University of Technology, Delft, The Netherlands, in 2024.
\end{IEEEbiography}

\vspace{11pt}

\begin{IEEEbiography}[{\includegraphics[width=1in,height=1.25in,clip,keepaspectratio]{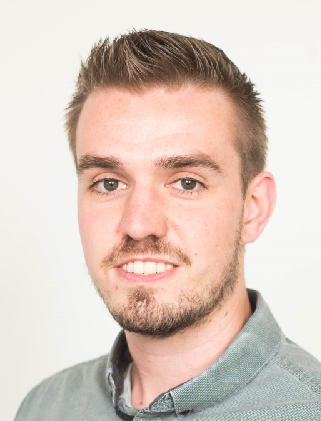}}]{Luke F. van Eijk}
received the B.Sc. (cum laude) and
M.Sc. (Hons.) degrees in mechanical engineering
from Eindhoven University of Technology, Eindhoven,
The Netherlands, in 2018 and 2021, with
a focus on dynamics and control. He is currently
pursuing the Ph.D. degree within the Department
of Precision and Microsystems Engineering, Delft
University of Technology, Delft, The Netherlands.

He is working as a Mechatronics Engineer at
ASMPT, Beuningen, The Netherlands. His research
interests are in the analysis and design of (non)linear
feedback controllers, with a particular focus on reset control and the hybrid
integrator-gain system.
\end{IEEEbiography}

\vspace{11pt}

\begin{IEEEbiography}[{\includegraphics[width=1in,height=1.25in,clip,keepaspectratio]{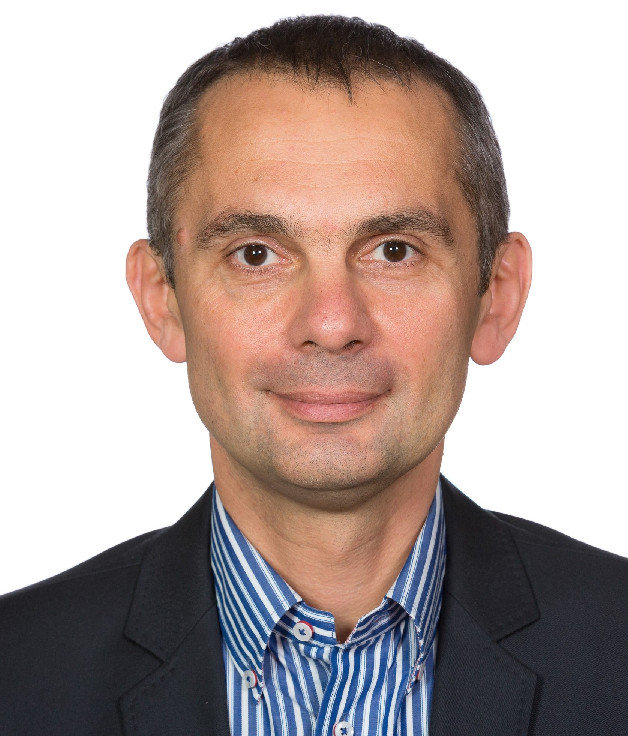}}]{Dragan Kosti\'c}
received the Ph.D. degree in control technology and robotics from the Eindhoven University of Technology, Eindhoven, The Netherlands, in 2004.

His professional positions range from research and teaching at knowledge institutions to professional engineering in commercial companies. Multidisciplinary system modelling and identification, data-based controls, and nonlinear control designs are his main areas of expertise. He works at ASMPT in Beuningen as the Research and Development director for mechatronics. His current research interests include modelling, dynamical analysis, and control of hi-tech mechatronic systems for semiconductor manufacturing.
\end{IEEEbiography}

\vspace{11pt}

\begin{IEEEbiography}[{\includegraphics[width=1in,height=1.25in,clip,keepaspectratio]{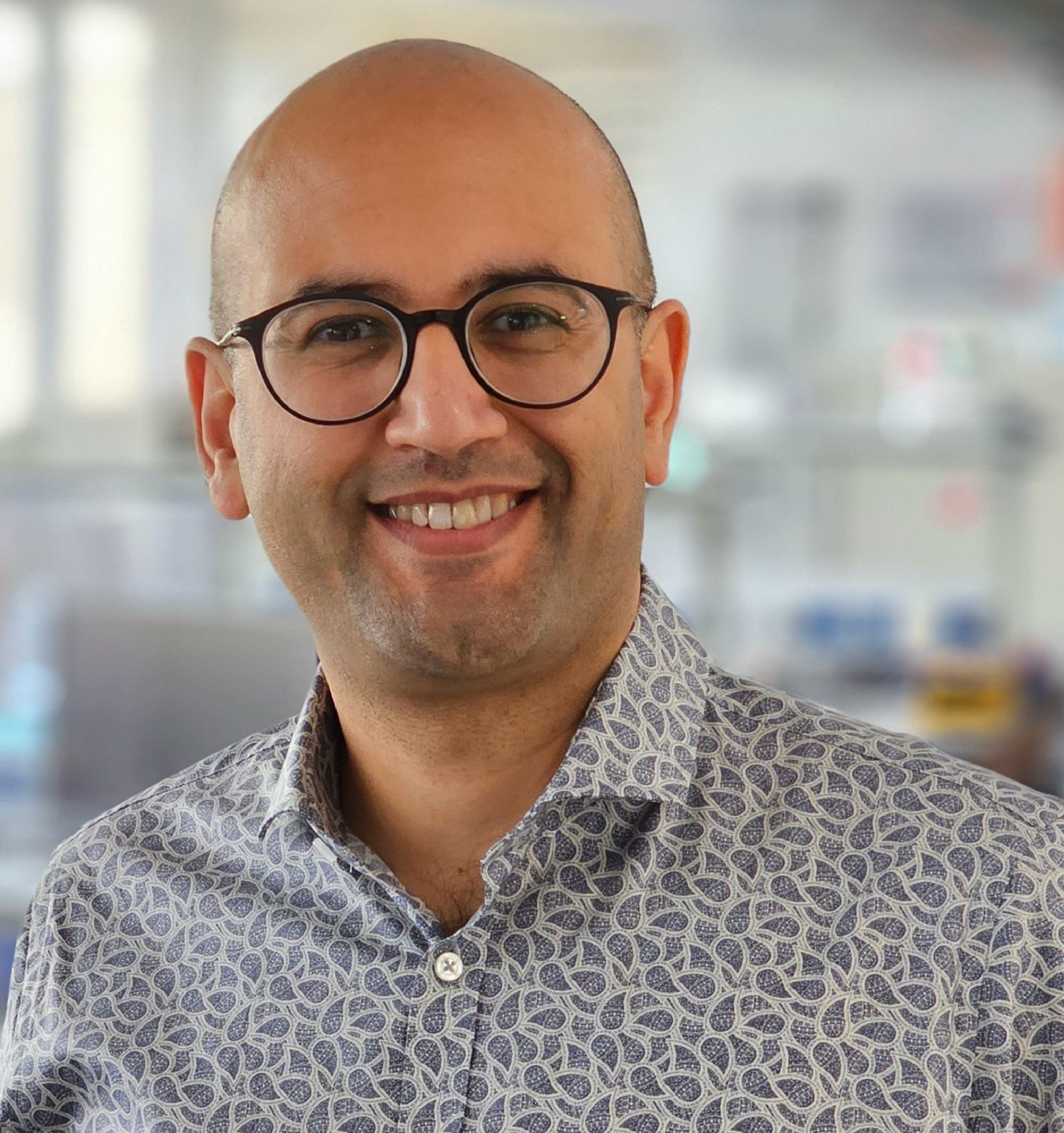}}]{S. Hassan HosseinNia} (Senior Member, IEEE) received the Ph.D. degree (Hons.) (cum laude) in
electrical engineering specializing in automatic control: application in mechatronics from the University of Extremadura, Badajoz, Spain, in 2013. He
has an industrial background, having worked with
ABB, Sweden. Since October 2014, he has been
appointed as a Faculty Member with the Department
of Precision and Microsystems Engineering, Delft
University of Technology, Delft, The Netherlands.
He has co-authored numerous articles in respected
journals, conference proceedings, and book chapters. His main research interests include precision mechatronic system design, precision motion control,
and mechatronic systems with distributed actuation and sensing.
Dr. HosseinNia served as the General Chair of the 7th IEEE International
Conference on Control, Mechatronics, and Automation (ICCMA 2019).
Currently, he is an editorial board member of “Fractional Calculus and Applied
Analysis,” “Frontiers in Control Engineering,” and “International Journal of
Advanced Robotic Systems (SAGE).”
\end{IEEEbiography}

\vspace{11pt}

\vfill

\end{document}